\documentclass[twocolumn,trackchanges]{aastex62}
\usepackage{graphicx}
\shorttitle{Faint satellites of Centaurus~A}
\shortauthors{Crnojevi\'c et al.}

\begin{document}


\title{The faint end of the Centaurus A satellite luminosity function}

\correspondingauthor{D. Crnojevi\'c}
\email{dcrnojevic@ut.edu}

\author{D. Crnojevi\'c}
\affil{University of Tampa, 401 West Kennedy Boulevard, Tampa, FL 33606, USA}
\affil{Department of Physics \& Astronomy, Texas Tech University, Box 41051, Lubbock, TX 79409-1051, USA}

\author{D. J. Sand}
\affiliation{Department of Astronomy and Steward Observatory, University of Arizona, 933 N Cherry Ave, Tucson, AZ 85719, USA}

\author{P. Bennet}
\affiliation{Department of Physics \& Astronomy, Texas Tech University, Box 41051, Lubbock, TX 79409-1051, USA}

\author{S. Pasetto}
\affiliation{Observatories of the Carnegie Institution for Science, 813 Santa Barbara Street, Pasadena, CA 91101, USA}
\affiliation{University of Tampa, 401 West Kennedy Boulevard, Tampa, FL 33606, USA}

\author{K. Spekkens}
\affiliation{Department of Physics and Space Science, Royal Military College of Canada, Box 17000, Station Forces, Kingston, ON K7L 7B4, Canada}

\author{N. Caldwell}
\affiliation{Harvard-Smithsonian Center for Astrophysics, Cambridge, MA 02138, USA}

\author{P. Guhathakurta}
\affiliation{Department of Astronomy \& Astrophysics, UCO/Lick Observatory, University of California Santa Cruz, 1156 High Street, Santa Cruz, CA 95064, USA}

\author{B. McLeod}
\affiliation{Harvard-Smithsonian Center for Astrophysics, Cambridge, MA 02138, USA}

\author{A. Seth}
\affiliation{Department of Physics and Astronomy, University of Utah, Salt Lake City, UT 84112, USA}

\author{J. D. Simon}
\affiliation{Observatories of the Carnegie Institution for Science, 813 Santa Barbara Street, Pasadena, CA 91101, USA}

\author{J. Strader}
\affiliation{Department of Physics and Astronomy, Michigan State University, East Lansing, MI 48824, USA}

\author{E. Toloba}
\affiliation{Department of Physics, University of the Pacific, 3601 Pacific Avenue, Stockton, CA 95211, USA}



\begin{abstract}

The Panoramic Imaging Survey of Centaurus and Sculptor (PISCeS) is 
constructing a wide-field map of the resolved stellar populations in
the extended halos of these two nearby, prominent galaxies. We present
new Magellan/Megacam imaging of a $\sim3$~deg$^2$ area around
Centaurus~A (Cen~A), which filled in much of our coverage to its
south, leaving a nearly complete halo map out to a projected radius of $\sim$150~kpc and
allowing us to identify two new resolved dwarf galaxies. We have
additionally obtained deep Hubble Space Telescope (HST) optical
imaging of eleven out of the thirteen candidate dwarf galaxies
identified around Cen~A and presented in \citet{crnojevic16}: seven
are confirmed to be satellites of Cen~A, while four are found to be background
galaxies. We derive accurate distances, structural parameters,
luminosities and photometric metallicities for the seven candidates
confirmed by our HST/ACS imaging. We further study the stellar
population along the $\sim$60~kpc long (in projection) stream associated with Dw3,
which likely had an initial brightness of $M_{V}$$\sim$$-$15 and shows
evidence for a metallicity gradient along its length.  Using the total
sample of eleven dwarf satellites discovered by the PISCeS survey, as
well as thirteen brighter previously known satellites of Cen~A, we
present a revised galaxy luminosity function for the Cen~A group down
to a limiting magnitude of $M_V\sim-8$, which has a slope of
$-1.14\pm0.17$, comparable to that seen in the Local Group and in
other nearby groups of galaxies.  

\end{abstract}

\keywords{galaxies: dwarf  --- galaxies: evolution ---
  galaxies: halos --- groups: individual (CenA) --- galaxies: luminosity function --- galaxies: photometry
}


\section{Introduction}

Observations on large scales ($\gtrsim$10 Mpc) are consistent with a
Universe dominated by dark energy and Cold Dark Matter, along with a
small baryonic component \citep[e.g.,][]{Planck16}.  Within this
$\Lambda$+Cold Dark Matter ($\Lambda$CDM) model for structure
formation, galaxies grow hierarchically within dark matter halos
\citep[e.g.,][]{Springel06}, and many detailed galaxy properties are
now reproduced in dark matter simulations that include the effects of
baryonic physics \citep[e.g.,][]{Vogel14}.  However, on scales
comparable to and below the size of individual galaxy halos
($\lesssim$1 Mpc), significant challenges to the $\Lambda$CDM
framework have been raised \citep[for a recent review, see
][]{Bullock17}, including the `missing satellites problem'
\citep{moore99,klypin99}, the `too big to fail' problem
\citep{boylan11}, and the apparent planes of satellites around nearby
galaxies \citep[e.g.,][]{Pawlowski12,Pawlowski13,ibata13,Muller18}.

Significant progress has been made in addressing the small-scale
challenges to the $\Lambda$CDM paradigm on both the theoretical and
observational fronts.  Numerical simulations that include a
sophisticated treatment of baryonic physics show improved comparisons
with dwarf galaxies in the Local Group
\citep[e.g.,][]{Brooks13,Sawala16,Wetzel16}, while a critical
assessment of the completeness limit of current searches for dwarf
galaxies around the Milky Way (MW) indicate that the 'missing
satellites problem' is not as severe as initially thought
\citep{koposov08, tollerud08,hargis14, Kim17}.  Meanwhile, further
Local Group dwarf galaxy discoveries \citep[e.g. most
recently][]{Drlica16,Torrealba18,Koposov18} add to the current total
and point to a rich bounty of new satellites in the era of the Large
Synoptic Survey Telescope.

To fully test the $\Lambda$CDM paradigm on sub-galactic scales,
however, we must also look beyond the Local Group to measure dwarf and
other substructure properties around primary halos with different
masses, morphologies and environments.  Recent progress has been made
in several nearby systems using deep, wide-field imaging
\citep[e.g.,][]{chiboucas09,Sand14,Sand15,crnojevic14b,crnojevic16,carlin16,toloba16,bennet17,Carrillo17,Danieli17,Smercina17,Smercina18},
as well as wide-field spectroscopy \citep{Geha17}.  Searches for
isolated dwarf galaxies provide further constraints on reionization
effects and on galaxy formation mechanisms, for instance in the absence
of tidal and ram pressure stripping
\citep{sand15b,tollerud15,Janesh17,tollerud18}.

Centaurus~A (Cen~A) is the closest accessible elliptical galaxy, and
it is the central galaxy of a relatively rich group
\citep[e.g.,][]{kara07}. We have thus chosen Cen~A as one of the
targets of our Panoramic Imaging Survey of Centaurus and Sculptor
(PISCeS), a wide-field imaging survey using the Megacam imager at the
Magellan Clay telescope.  One of the principal goals of PISCeS is to
identify new, faint dwarf galaxies around Cen~A and around the spiral
Sculptor (NGC~253; located in a loose group of galaxies) by imaging
their resolved stellar populations and compare the properties of these
dwarfs to those of Local Group and simulated dwarfs.
In previous work around Cen~A, we have highlighted a pair of faint
satellites at $D$$\approx$90 kpc in projection \citep{crnojevic14b},
and have presented a comprehensive red giant branch (RGB) star map of
Cen~A's halo, highlighting new streams, dwarfs galaxies, and other halo
substructures \citep{crnojevic16}. Our work represents the most
complete census of the halo stellar populations and the satellites
within $\sim150$~kpc of Cen~A. Here we present two new dwarf galaxy
candidates from our 2017 observing season, which focused on extending
the spatial coverage of the survey to the south of Cen~A.  We also present 
Hubble Space Telescope (HST) follow-up of 11 dwarf galaxy
candidates, along with a detailed look at a disrupting dwarf galaxy
and its associated stellar stream. HST follow-up of our ground-based
discoveries allows for improved distance, structural parameters, and luminosity
measurements, and, in some cases, is necessary for determining whether
the PISCeS ground-based candidates are indeed dwarf galaxies at the
distance of Cen~A. Based on the results from these HST data and
our continuing ground-based campaign,  we provide a preliminary
estimate of the dwarf galaxy luminosity function (LF) of Cen~A to compare
with those calculated for other Local Volume groups of galaxies.
Throughout this work, we assume a Cen~A distance of $D$=3.8~Mpc \citep{harrisg10}.


\section{The PISCeS survey} \label{sec:pisces}

We begin by briefly describing the PISCeS survey as it pertains to Cen~A.  For more details on  our survey strategy and observational methods, see \citet{crnojevic14b,crnojevic16}, and for our preliminary results around the nearby spiral galaxy NGC~253, see \citet{Sand14} and \citet{toloba16}.  

The ultimate observational goal of PISCeS is to image the halos of
Cen~A and NGC~253 in the Sculptor group out to a projected radius of
$D$$\sim$150 kpc, deep enough to resolve $\sim$1--2 mag below the tip
of the red giant branch (limiting magnitudes: $g,~r$$\approx$26-26.5
mag). The data of this areal coverage is comparable to that of the
Pan-Andromeda Archaeological Survey of M31
\citep[e.g.,][]{mcconnachie09}, though three magnitudes less deep in
absolute magnitude. This allows for direct comparison of the satellite
and substructure properties between the Cen~A and M31 systems, and extends our general knowledge of substructure to new systems and environments.  

To image the outer halo of Cen~A, we mosaic individual pointings of
the Megacam imager \citep{mcleod15} on the Magellan Clay telescope.
Megacam has a $\sim$24'$\times$24' field of view, and a typical
pointing is observed for 6$\times$300 s in each of the $g$ and $r$
bands.  The data are reduced in a standard way  by the Smithsonian
Astrophysical Observatory Telescope Data Center \citep[see][for
further details]{mcleod15,crnojevic16}, and point-spread function
(PSF) photometry is performed on the stacked final images using the
{\sc daophot} and {\sc allframe} software suite
\citep{stetson87,stetson94}.  Instrumental magnitudes are calibrated
to the Sloan Digital Sky Survey (SDSS) system using standard star
field observations obtained on photometric nights; calibration of the
overall survey is facilitated by small pointing overlaps between
adjacent fields (typically $\sim$2 arcmin).  Artificial star tests are
run on all images to quantify completeness and magnitude
uncertainties.

\begin{figure*}
 \centering
\includegraphics[width=9cm]{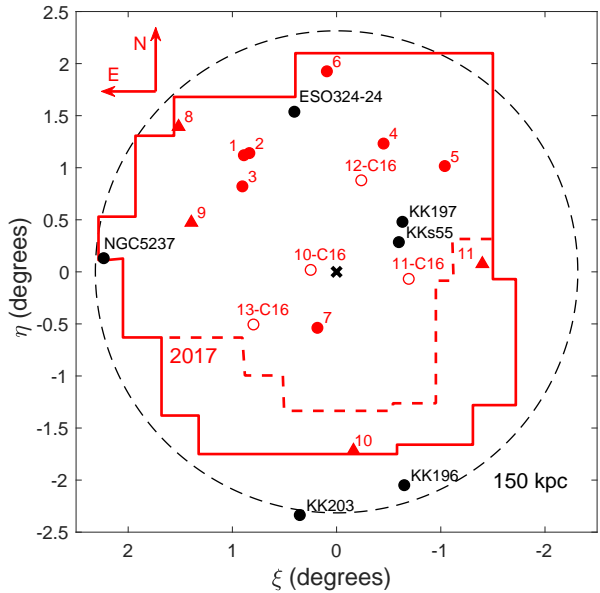}
\caption{Footprint of the PISCeS survey to date (red polygon) in
  standard coordinates centered on Cen~A (black cross); the area
  surveyed in our latest observing run (2017; i.e., the area not
  reported in \citealt{crnojevic16}) is to the south of the red dashed
  line; the 150~kpc projected galactocentric radius is shown as a
  black dashed circle. Cen~A dwarfs that were known prior to the
  PISCeS survey are plotted as filled black circles; confirmed
  satellites discovered in PISCeS are filled red circles or triangles
  (numbered following their names, e.g., 1 is for CenA-MM-Dw1), where
  the triangles indicate dwarfs that have not been observed with HST;
  open red circles are candidate satellites from \citet{crnojevic16} that turned out to be background objects as deduced from HST imaging (see Sect.~\ref{sec:pisces} for the nomenclature convention).}
\label{map}
\end{figure*}

The status of the Cen~A PISCeS program as of 2017 can be seen in
Figure~\ref{map}.  The data obtained prior to 2017 were published in
\citet{crnojevic16}, where 13 new dwarf candidates were presented
(along with other halo substructures).  These dwarf candidates span an
absolute magnitude range of $-$7.2 $>$ $M_{V}$ $>$ $-$13.0 and
half-light radius range of $\sim$220--2900 pc, shedding light on Cen~A's faint satellite population ($M_V\gtrsim-10$) for the first time.  Here we present HST follow-up imaging of 11 of these dwarf galaxy candidates to confirm their status as Cen~A dwarfs and to precisely measure their physical properties.  We also present HST pointings along the dramatic tidal stream associated with Dw3.  In the next section, we present further results of our 2017 ground-based Megacam campaign, focused on filling in our southern coverage of Cen~A's halo, where we discovered two new dwarf galaxies that are discussed here for the first time.

We have obtained data in 2018 to complete our survey out to a galactocentric radius of $\sim150$~kpc. The data reduction for that dataset is still in progress and will be presented elsewhere; however a visual inspection of the images does not reveal any new candidate dwarf satellite in the remaining surveyed area. 

\section{The Magellan 2017 dataset: Two New Cen~A Dwarfs}\label{sec:newdwarfs}

We continued the PISCeS campaign around Cen~A in 2017, observing with
Megacam/Magellan Clay on the nights of 2017 April 20-24 (UT).  During
the span of these 5 nights, we collected data for 20 new fields,
focused on filling in areas south of Cen~A, as can be seen by the
dashed region in Figure~\ref{map} (which shows a total of 95 pointings
obtained up to the 2017 campaign). The data for the 20 new Megacam
fields were generally obtained in photometric conditions, with the
seeing ranging from 0.5--0.9~arcsec ($g$ band).  The data were reduced as described in \citet{crnojevic16} and Section~\ref{sec:pisces}.

\begin{figure*}
 \centering
\includegraphics[width=10cm]{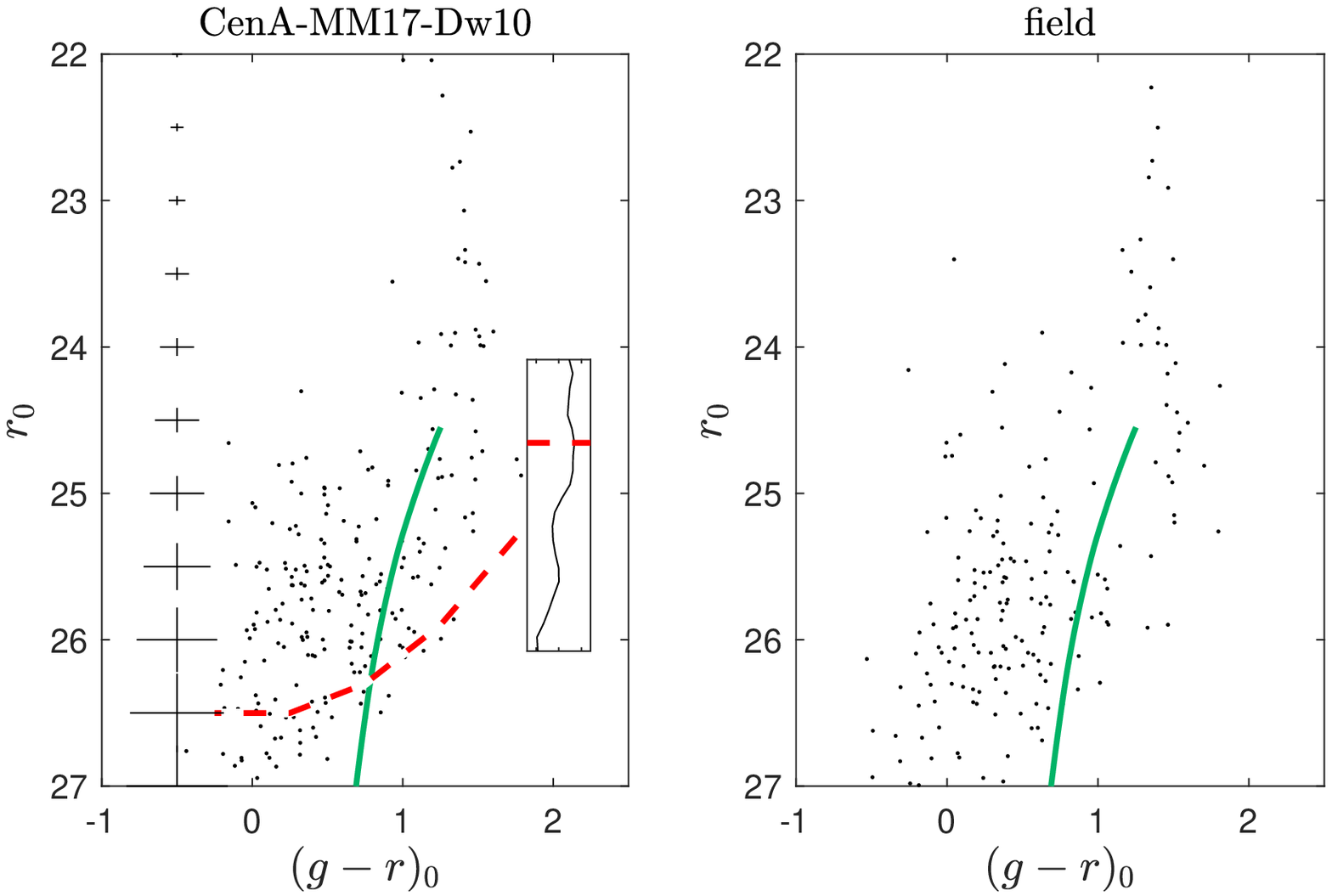}
\includegraphics[width=7cm]{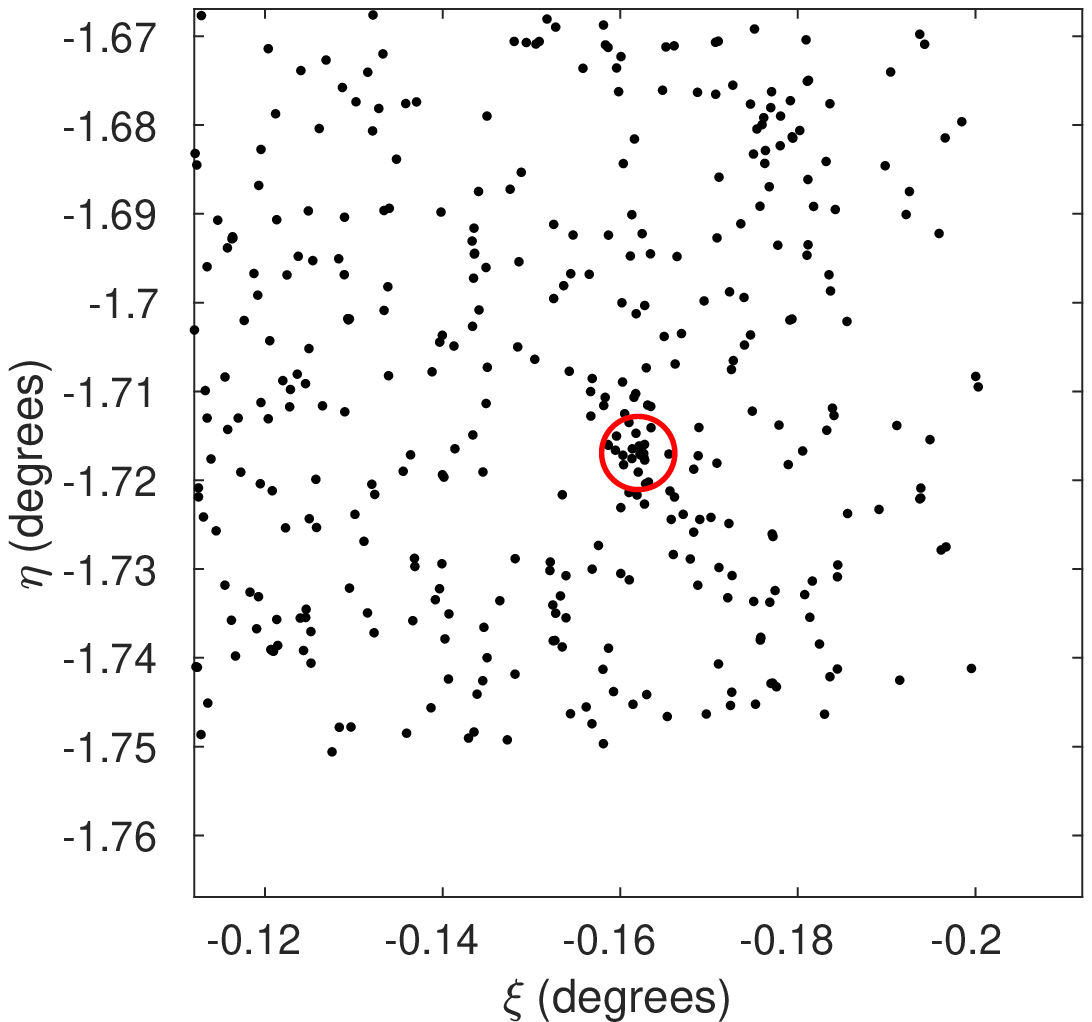}
\includegraphics[width=10cm]{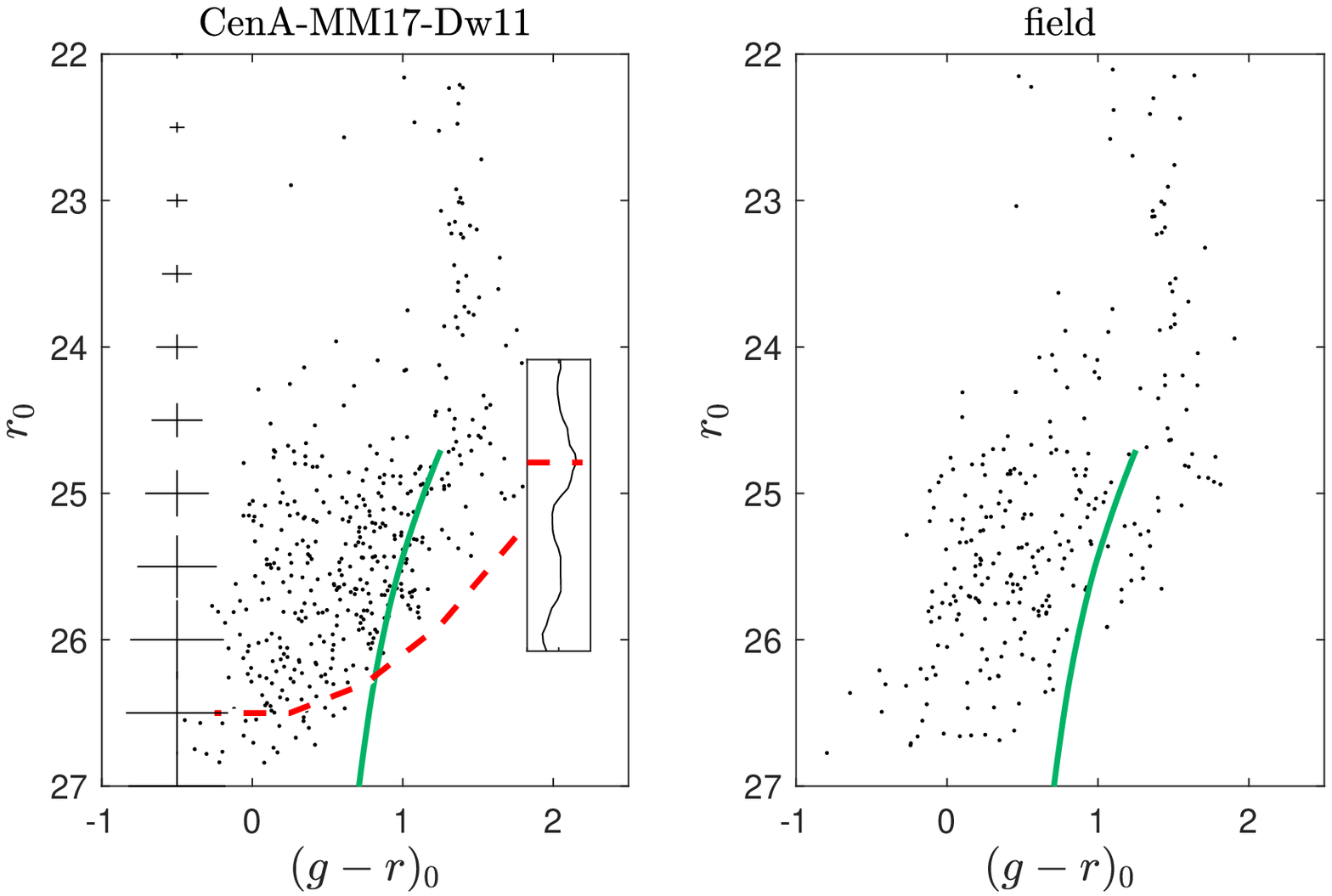}
\includegraphics[width=7cm]{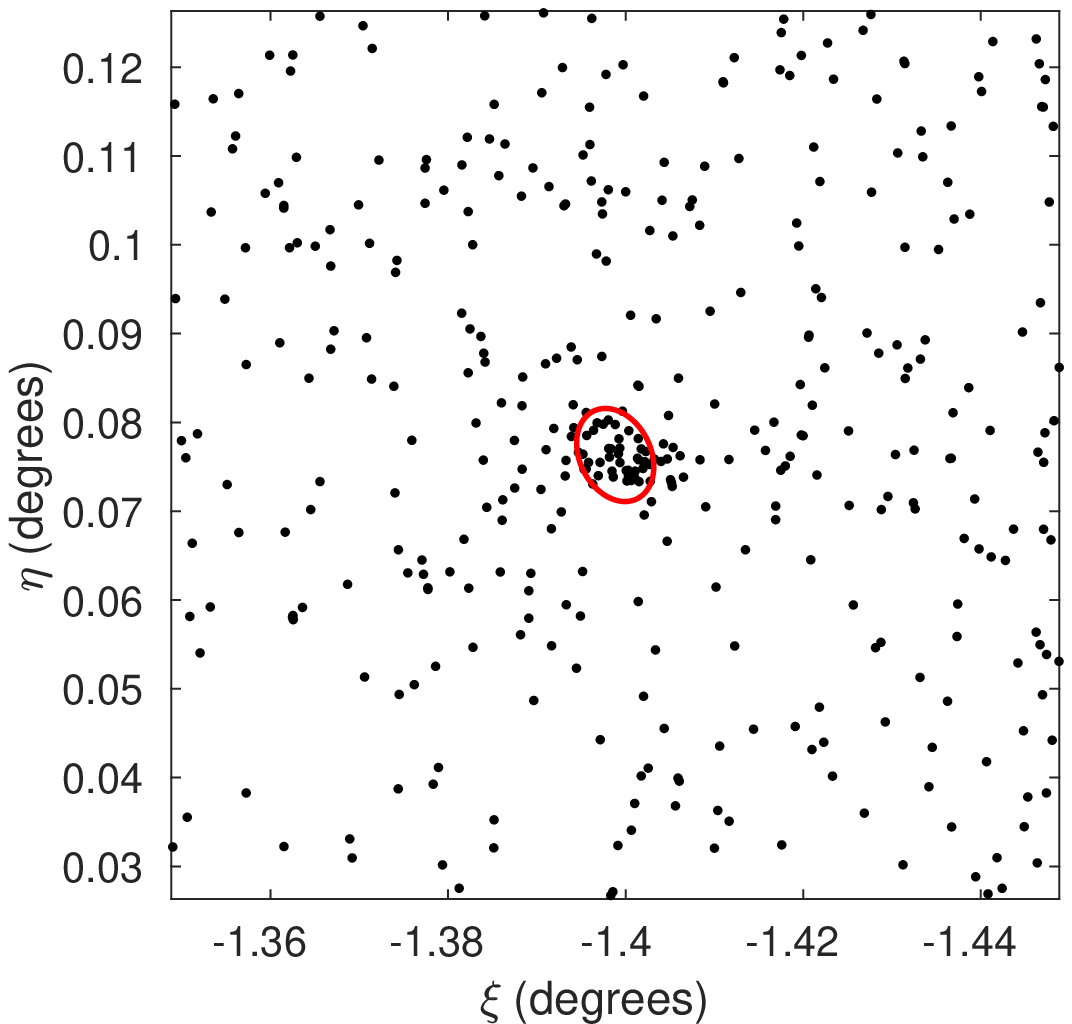}
\caption{Dereddened Magellan/Megacam CMDs for CenA-MM17-Dw10 (top
  panels) and CenA-MM17-Dw11 (bottom panels). We plot stars within a
  box of $0.6\times0.6$~arcmin$^2$ and $0.9\times0.9$~arcmin$^2$
  centered on Dw10 and Dw11, respectively. Isochrones with a
  12~Gyr age and a metallicity [Fe/H]$=-1.5$ \citep{dotter08} are
  shifted to the measured distance of each dwarf. The red dashed line
  indicates the $50\%$ completeness level, and photometric errors as
  derived from artificial star tests are shown on the left side of the
  CMD. The inset plots in the left panels show the LF
  after convolution with a Sobel filter, and the derived TRGB
  magnitude (red dashed line). A background field CMD drawn from the
  Magellan pointing containing the dwarf and rescaled in area is shown for
  comparison in the middle panels. In the right panels, we show a
  $3\times3$~arcmin$^2$ cutout of the dwarf RGB stars' spatial
  distribution in the Magellan/Megacam pointings,
  centered on the dwarfs. The red circles indicate each dwarf's
  measured half-light radius.}
\label{dw1011}
\end{figure*}

We found two new dwarf galaxy candidates in the 2017 Magellan dataset, which are marked in Figure~\ref{map}. Since some of our dwarf candidates from \citet{crnojevic16} turned out to be background objects (see Sect.~\ref{confirm_sec}), we rename those Dw10-C16 to Dw13-C16 (from the original \citealt{crnojevic16} nomenclature), and we dub the two new discoveries CenA-MM17-Dw10 and CenA-MM17-Dw11, or to follow the original nomenclature, Dw10 and Dw11.
These dwarfs were first found via visual inspection and then confirmed
based on the red giant branch (RGB) map of stars consistent with the distance to Cen~A in each field, where they stand out as clear overdensities.  We show the dereddened color magnitude diagram (CMD) and RGB spatial map of each dwarf in Figure~\ref{dw1011}.  Both objects are faint, but clearly detected above the background.  Each consists of an old stellar population (the isochrones shown in Figure~\ref{dw1011} are of a 12 Gyr old, [Fe/H]=$-$1.5 stellar population; \citealt{dotter08}) with no signs of recent star formation.

We measure the distance to Dw10 and Dw11 as described in
\citet{crnojevic16}.  We use the standard tip of the red giant branch
(TRGB) method \citep[e.g.,][and see further discussion in
Section~\ref{dist_sec} in relation to our HST data]{dacosta90,lee93}, measuring
a discontinuity in its LF with a Sobel edge detection filter. The newly discovered dwarfs, Dw10 and Dw11, have TRGB distances consistent with Cen~A (Table~\ref{dw_mag}), confirming their association.  In projection, Dw10 and Dw11 are $\sim110$ and $\sim90$~kpc (or 1.72 and 1.40~deg) from Cen~A, respectively.

To measure the structural parameters and luminosities of Dw10 and
Dw11, we use the method of moments as presented in
\citet{crnojevic14a}.  First, the surface brightness profile of each
dwarf is found by summing the area-normalized flux of stars within an
RGB selection box as a function of radius, correcting for field
contamination and incompleteness based on our artificial star tests.
In order to correct for unresolved light from stars below our
detection limit, we directly measure the image flux within a central
aperture and rescale the surface brightness profile to match it.  This
rescaled surface brightness profile is fit to an exponential  using
least-squares minimization in order to measure the half-light radius
and central surface brightness.  Finally, the absolute magnitude is
computed by integrating the best-fit exponential profile.  The final
derived quantities for Dw10 and Dw11 are presented in
Table~\ref{dw_mag}.  The properties of these two new dwarfs are broadly consistent with the dwarf population found in our earlier work \citep{crnojevic16}.

We also searched the HI Parkes All Sky Survey
\citep[HIPASS;][]{barnes01} for neutral gas at the position of the two
new dwarfs, and present their 5-$\sigma$ upper limits in
Table~\ref{dw_mag}.  As with the HI limits of the other Cen~A dwarfs
found by PISCeS \citep{crnojevic16}, the HIPASS data are not sensitive
enough to confirm or exclude the presence of HI in these dwarfs below
$3\times10^6 M_\odot$.

\tabletypesize{\scriptsize}
\begin{deluxetable} {lcc}
\tablecolumns{3}
\tablecaption{Properties of new PISCeS 2017 dwarfs}
 \tablehead{\colhead{Parameter}  & \colhead{Dw10} &\colhead{Dw11}}
\startdata
RA (h:m:s) & 13:24:32.9$\pm1$'' & 13:17:49.2$\pm1$''\\
Dec (d:m:s) & $-44$:44:07.1$\pm2$'' & $-42$:55:36.8$\pm8$'' \\
$E_{(B-V)}$ & $0.09$ & $0.12$ \\
$(m-M)_0$ (mag) & $27.57\pm0.29$ & $27.73\pm0.22$ \\
D (Mpc) & $3.27^{+0.41}_{-0.46}$& $3.52^{+0.33}_{-0.37}$\\
$D_{\rm CenA, proj}$ (deg) & $1.72$ & $1.40$ \\
$D_{\rm CenA, proj}$ (kpc) & $112$ & $91$ \\
$\epsilon$ & $<0.27$\tablenotemark{a} & $0.27\pm0.21$ \\
$r_{h}$ (arcmin) & $0.25\pm0.06$ & $0.33\pm0.04$ \\
$r_{h}$ (kpc) & $0.24\pm0.06$& $0.34\pm0.04$ \\ 
$\mu_{V,0}$ (mag~arcsec$^{-2}$) & $26.6\pm0.9$ & $25.8\pm0.4$ \\ 
$M_V$ (mag) &$-7.8\pm1.2$ &$-9.4\pm0.6$ \\
$L_*$ ($10^5 L_\odot$) &$1.1\pm1.4$ &$4.7\pm2.7$ \\
$M_{\rm HI}$\tablenotemark{b} ($10^6 M_\odot$) & $\lesssim4.0$ & $\lesssim3.1$ \\
$M_{\rm HI}/L_*$ ($M_\odot/L_\odot$) & $\lesssim36.7$ & $\lesssim6.6$ \\
\enddata
\tablenotetext{a}{Only an upper limit on the ellipticity, $\epsilon$, was measurable.}
\tablenotetext{b}{5~$\sigma$ upper limits from HIPASS.}
\label{dw_mag}
\end{deluxetable}

\section{HST Imaging: Data and photometry} \label{data}

Including the discovery of Dw10 and Dw11, PISCeS has uncovered fifteen
dwarf candidates around Cen~A (see Table~\ref{summary}). As mentioned
above, we obtained HST observations of eleven of the thirteen dwarf
candidates found in \citet{crnojevic16}, which are marked as red
circles (both filled and unfilled) in Figure~\ref{map}.

HST follow-up imaging was obtained with the Wide Field Channel (WFC) of the 
Advanced Camera for Surveys (ACS).
Most of the targets were observed as part of the 
program GO-13856 (PI: Crnojevi\'c), with the exception of 
CenA-MM-Dw3, which was observed as part of program GO-14259 
(PI: Crnojevi\'c); see Table~\ref{summary} for a summary. Each target was observed for a total of one 
orbit (two orbits for CenA-MM-Dw3) in the $F606W$ and $F814W$ filters, 
which broadly correspond to the Johnson-Cousins $V$ and $I$ bands
(exposure times of $\sim1100/2500$~sec per filter, for one and two
orbits, respectively).

Parallel observations were simultaneously obtained
with the Wide Field Camera 3 (WFC3) UVIS channel, with 
the same filters and similar exposure times. The parallel pointings 
serve as background/foreground fields (to clean the CMDs from contaminating
resolved objects), as well as control fields of the Cen~A
halo (to study the halo properties including possible stellar
population gradients).

\tabletypesize{\scriptsize}
\begin{deluxetable*} {lcccc}
\tablecolumns{5}
\tablecaption{Cen~A satellites discovered in PISCeS.}

 \tablehead{\colhead{Galaxy} & \colhead{Alternative name}  &
   \colhead{HST Program ID} &\colhead{Confirmed} &\colhead{Ref\tablenotemark{a}}}
\startdata
Dw1 & CenA-Dw-133013-415321 & 13856 & y & 1\\
Dw2 & CenA-Dw-132956-415220 & 13856 & y & 1\\
Dw3 & $-$ & 14259 & y & 2\\
Dw4 & CenA-Dw-132302-414705 & 13856 & y & 2\\
Dw5 & CenA-Dw-131952-415938 & 13856 & y & 2\\
Dw6 & CenA-Dw-132557-410538 & 13856 & y & 2\\
Dw7 & CenA-Dw-132628-433318 & 13856 & y & 2\\
Dw8 & $-$ & $-$ & y & 2\\
Dw9 & $-$ & $-$ & y & 2\\
Dw10-C16 & CenA-Dw-132649-430000 & 13856 & n & 2\\
Dw11-C16 & CenA-Dw-132140-430457 & 13856 & n & 2\\
Dw12-C16 & CenA-Dw-132410-420823 & 13856 & n & 2\\
Dw13-C16 & CenA-Dw-132951-433109 & 13856 & n & 2\\
Dw10 & CenA-MM17-Dw10 & $-$ & y & 3\\
Dw11 & CenA-MM17-Dw11 & $-$ & y & 3\\
\enddata

\tablenotetext{a}{References: 1=\citet{crnojevic14b}; 2=\citet{crnojevic16}; 3=This work.}
\label{summary}
\end{deluxetable*}

\subsection{Photometry} 

Point spread function (PSF) photometry was performed on the
pipeline-produced \emph{.flt} images with the latest version (2.0) 
of the dedicated photometric package DOLPHOT \citep{dolphin02}.
Generally, we adopt the input parameters suggested by the 
DOLPHOT User's Guide for each camera, including the corrections
for CTE losses which are substantial for both ACS and WFC3.
For the most crowded of our pointings (the one targeting
Dw7, i.e., the candidate satellite closest to the center of Cen~A in projection, at $\sim0.5$~deg),
we set the parameters $FitSky=3$ and $img\_{RAper}=10$ to improve
the sky-fitting procedure. The photometry is then culled
with the following criteria: the sum of the crowding parameters
in the two bands is $<1$ (or $<0.6$ for the crowded
photometry case), the squared sum of the sharpness parameters
in the two bands is $<0.075$, and the photometric errors
as derived by DOLPHOT are $\lesssim0.3$ in each band.

\begin{figure}
 \centering
\includegraphics[width=7cm]{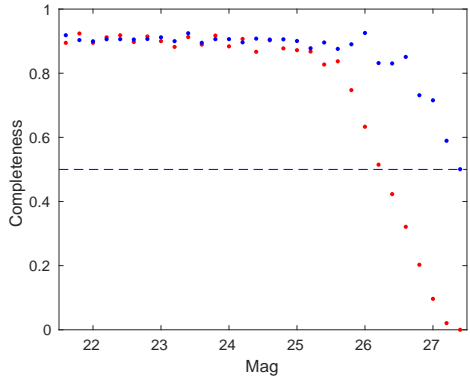}
\caption{Photometric completeness curves for Dw1 in the $F814W$ (red)
  and $F606W$ (blue) filters, as determined from our artificial star
  tests; these are representative for the overall sample of targets.
}
\label{compl}
\end{figure}

We subsequently perform artificial star tests in order to 
accurately assess photometric errors and incompleteness in the
HST data. The artificial stars are distributed evenly
both spatially and in color-magnitude space, and extend as faint
as two magnitudes below the faintest detected stars (after
quality cuts) to account for objects upscattered into the detectable
magnitude range due to blending and noise. For each field, we inject a number of artificial stars 
between a minimum of 200000, in order to ensure a robust statistics, 
and a maximum of 1200000, i.e., 10 times the number of sources
(after quality cuts) in the most crowded of our pointings (note that DOLPHOT adds one fake star at a time in order not to increase crowding artificially).
Photometry and quality cuts are performed in the same exact way as 
done for the original photometry; photometric errors are shown for each galaxy in the CMDs 
of the next section. Finally, representative completeness curves are shown in Fig.~\ref{compl} (the maximum completeness value is below $100\%$ because of spatial incompleteness).

\begin{deluxetable*}{lccc}
\tablecolumns{4}
\tablecaption{Properties of unresolved background galaxies.}
 \tablehead{\colhead{Parameter}  & \colhead{Dw10-C16} &\colhead{Dw11-C16} &\colhead{Dw12-C16}}
\startdata
RA (h:m:s) & 13:26:49.4$\pm1.52$'' & 13:21:40.3$\pm0.98$'' & 13:24:10.9$\pm1.00$'' \\
Dec (d:m:s) & $-43$:00:01$\pm1.46$'' & $-43$:05:00$\pm1.08$'' & $-42$:08:24$\pm9.00$'' \\
$m_{g}$ (mag) & 21.9$\pm$0.3 & 20.8$\pm$0.4 & 19.3$\pm$0.1  \\
$m_{r}$ (mag) & 21.3$\pm$0.4 & 19.6$\pm$0.5 & 18.8$\pm$0.2 \\
$r_{h}$ (arcsec) & 3.93$\pm$0.97 & 11.70$\pm$3.46 & 7.23$\pm$0.69 \\
$n$ (S\'ersic index) & 0.53$\pm$0.17 & 0.42$\pm$0.18 & 0.62$\pm$0.05\\
$\epsilon$ & 0.36$\pm$0.10 & 0.68$\pm$0.10 & 0.36$\pm$0.04 \\
\enddata
\end{deluxetable*}


\begin{figure}
 \centering
\includegraphics[width=8cm]{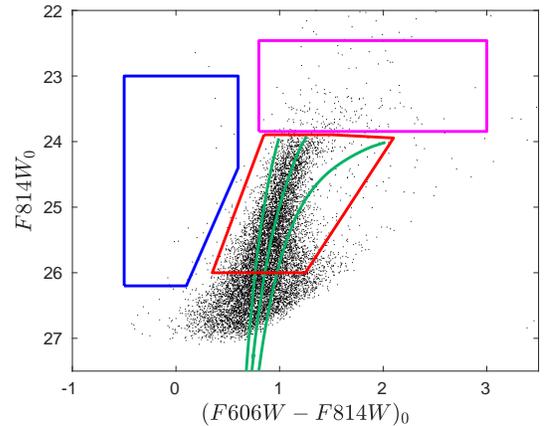}
\caption{Color-magnitude diagram of Dw1, identifying different stellar populations within the target galaxies: old red giant branch stars (red box; 10~Gyr isochrones with metallicities of [Fe/H]$=-2.0$, $-1.0$, and $-0.5$, from left to right); young massive stars ($\lesssim500$~Myr, blue box); and intermediate-age upper asymptotic giant branch stars (magenta box).  This CMD and the selection boxes are representative for our whole sample of targets.
}
\label{boxes}
\end{figure}

\subsection{Satellite confirmation and contaminants from HST imaging}  \label{confirm_sec}

\begin{figure*}
 \centering
\includegraphics[width=9.cm]{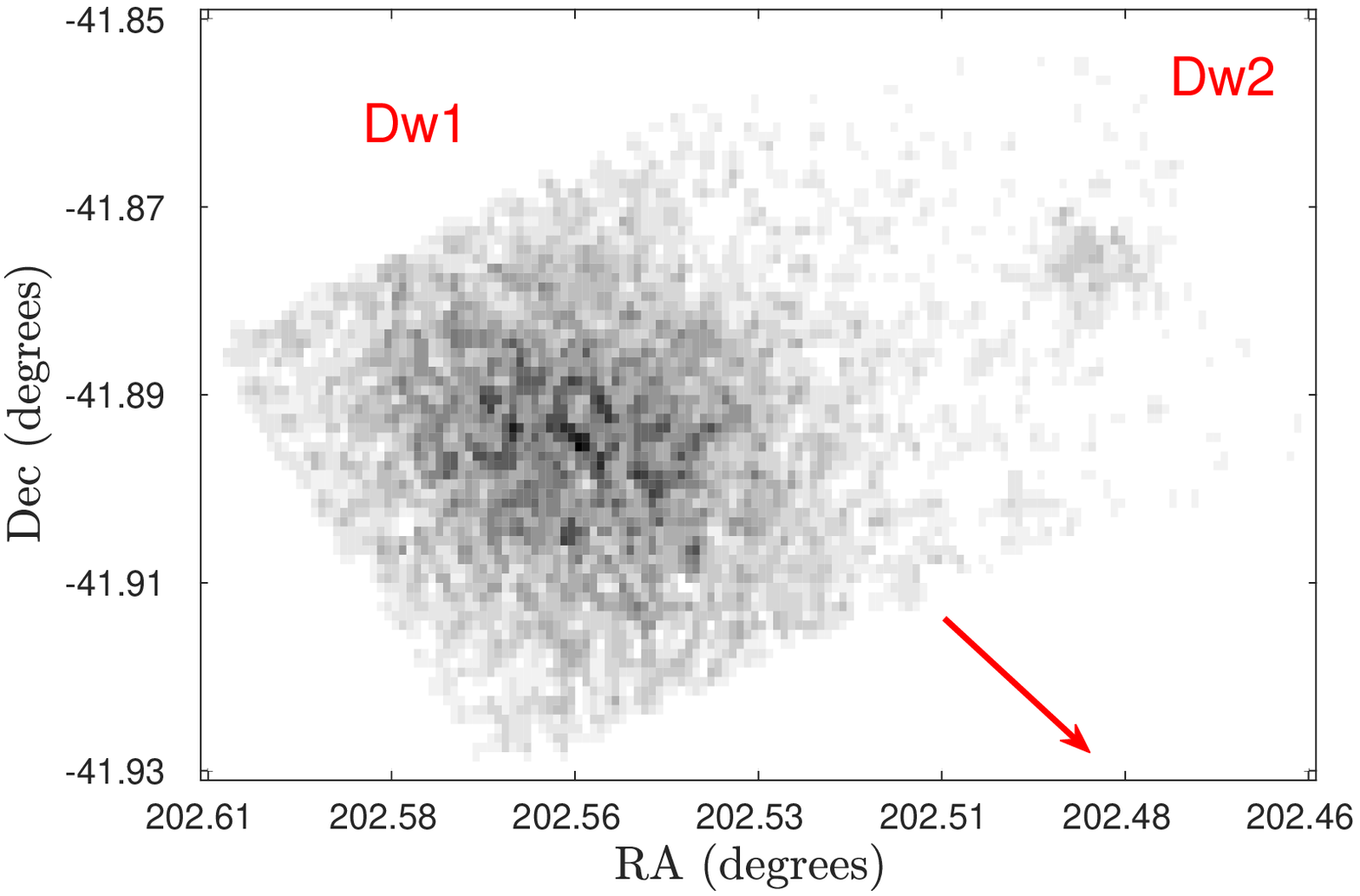}
\includegraphics[width=18cm]{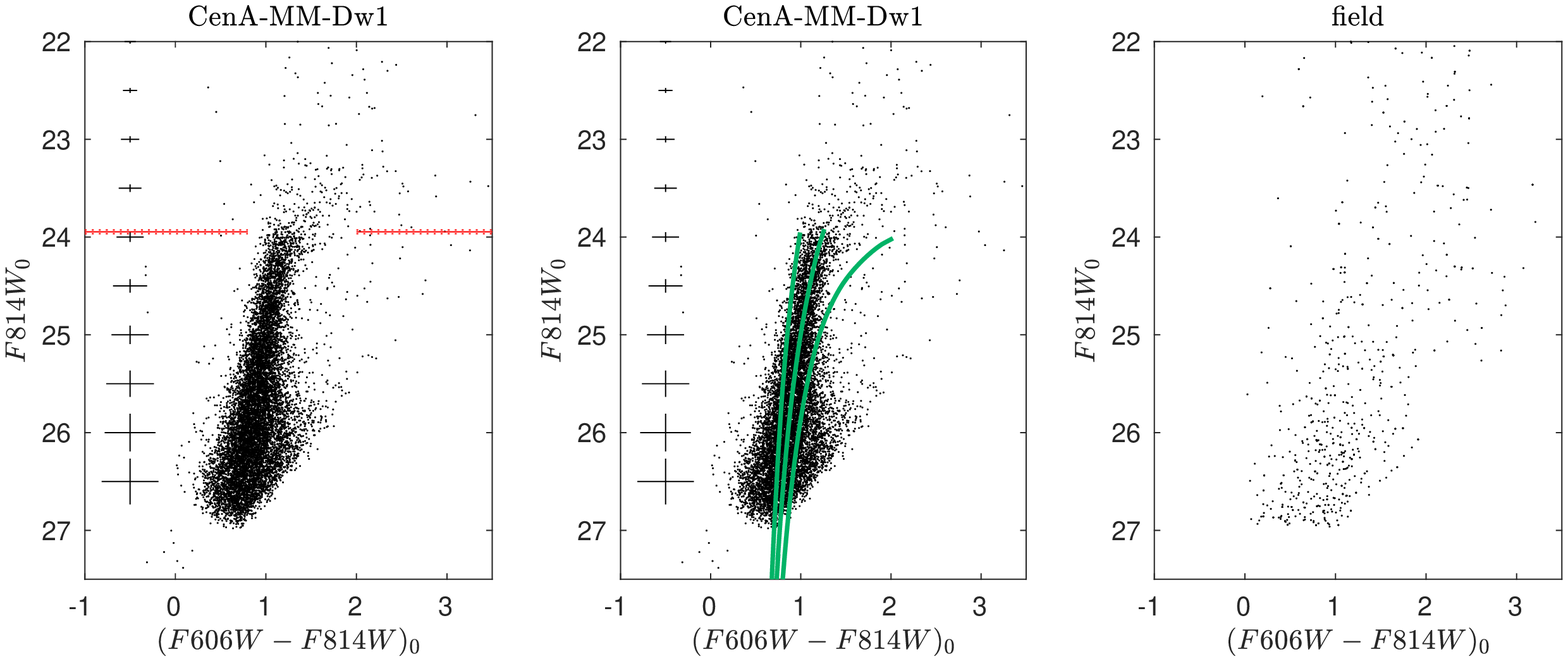}
\includegraphics[width=18cm]{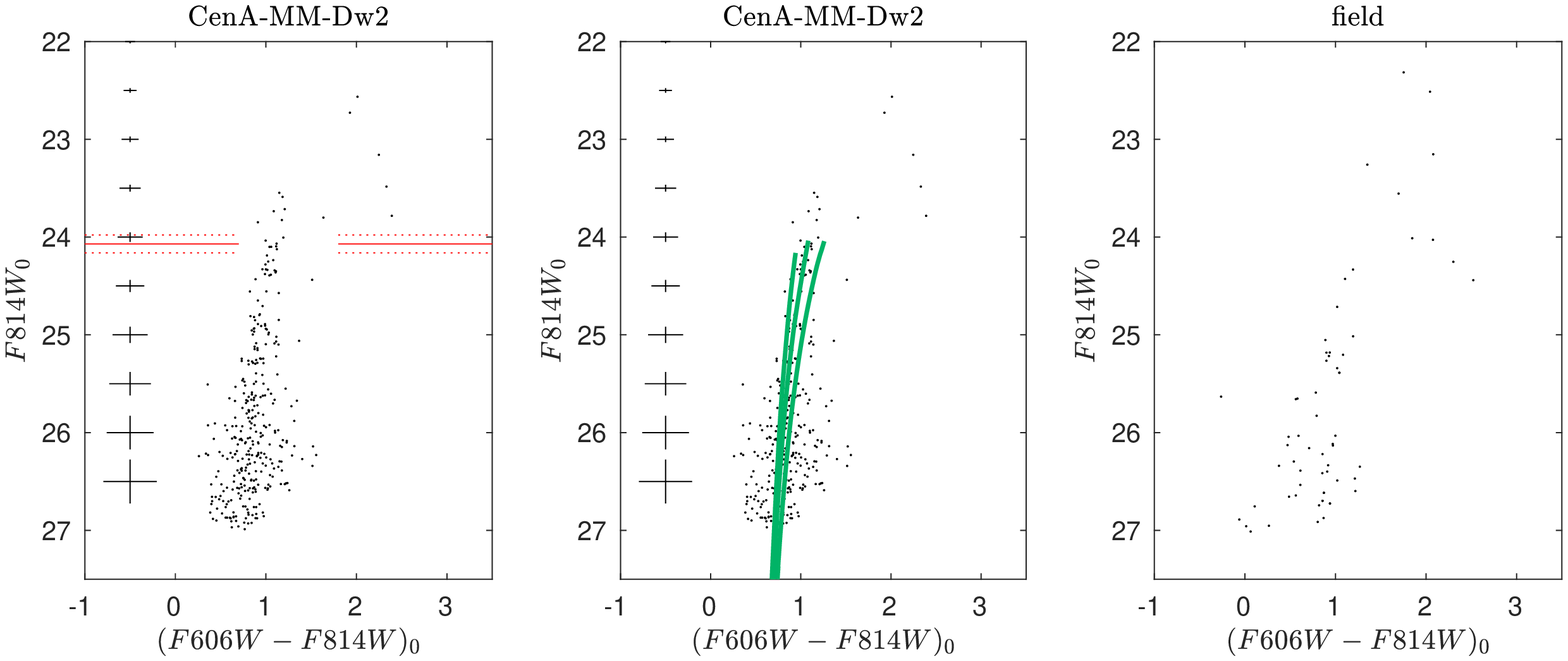}
\caption{\emph{Upper panel.} Stellar density map of RGB stars for the two ACS pointings targeting Dw1 and Dw2 (as labelled), derived for stars within the red selection box in Fig.~\ref{boxes}. The direction towards Cen~A is indicated by the red arrow. The former clearly overfills the ACS field of view. \emph{Central and bottom panels.} CMDs of Dw1 and Dw2 including stellar sources within $0.5 r_h$ and $r_h$, respectively; we also report photometric errors as derived from artificial star tests. In the \emph{left} panels, we draw the TRGB magnitude and the relative uncertainties (red lines and red dotted lines); in the \emph{central} panels, we overplot 10~Gyr isochrones with metallicities of [Fe/H]$=-2.0$, $-1.0$, and $-0.5$ for Dw1, and $-2.5$, $-1.5$, and $-1.0$ for Dw2 (green lines; \citealt{dotter08}); in the \emph{right panels}, an area-scaled field CMD is shown.
}
\label{dwonetwo}
\end{figure*}

\begin{figure*}
 \centering
\includegraphics[width=7.cm]{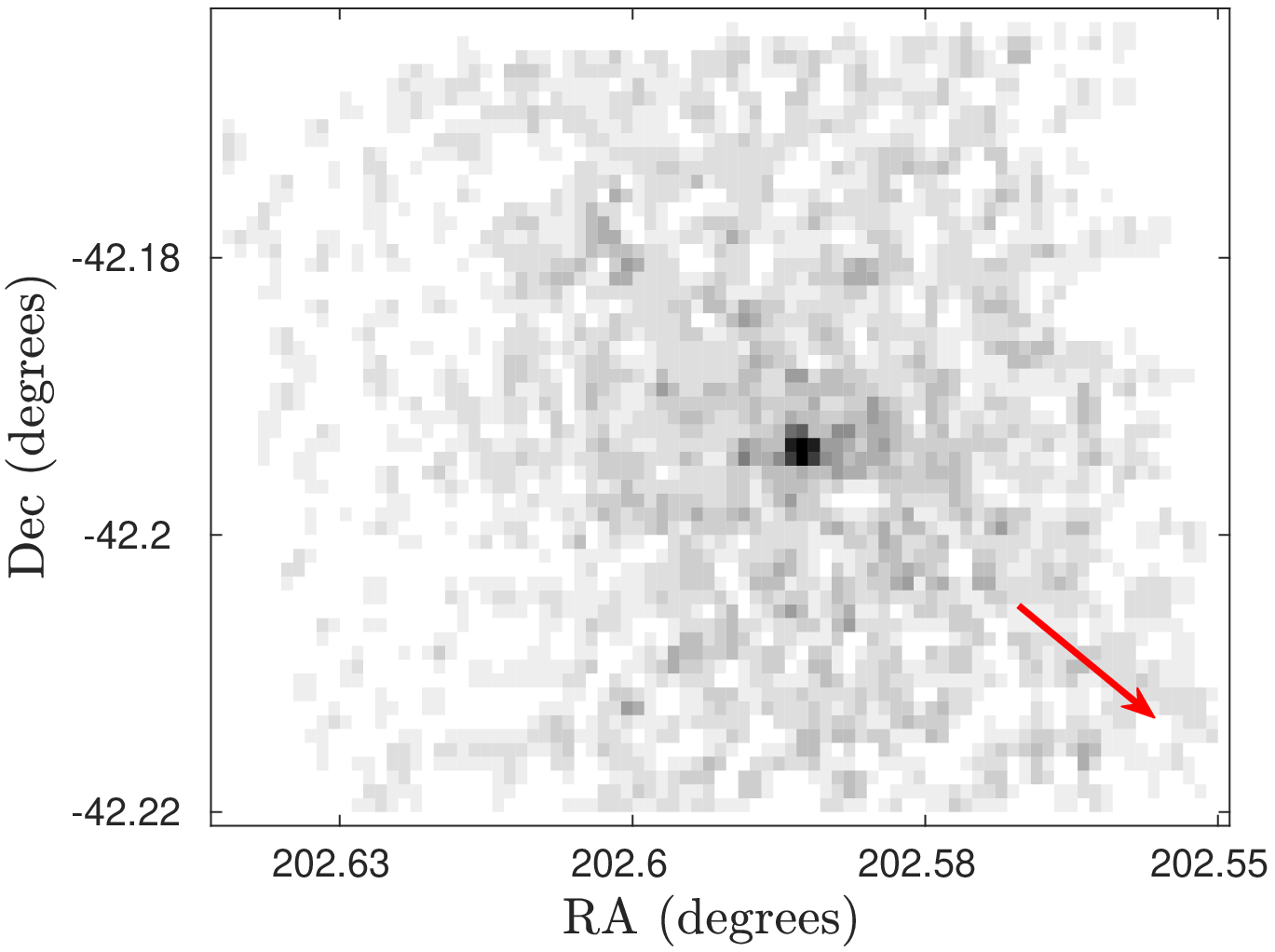}
\includegraphics[width=18.cm]{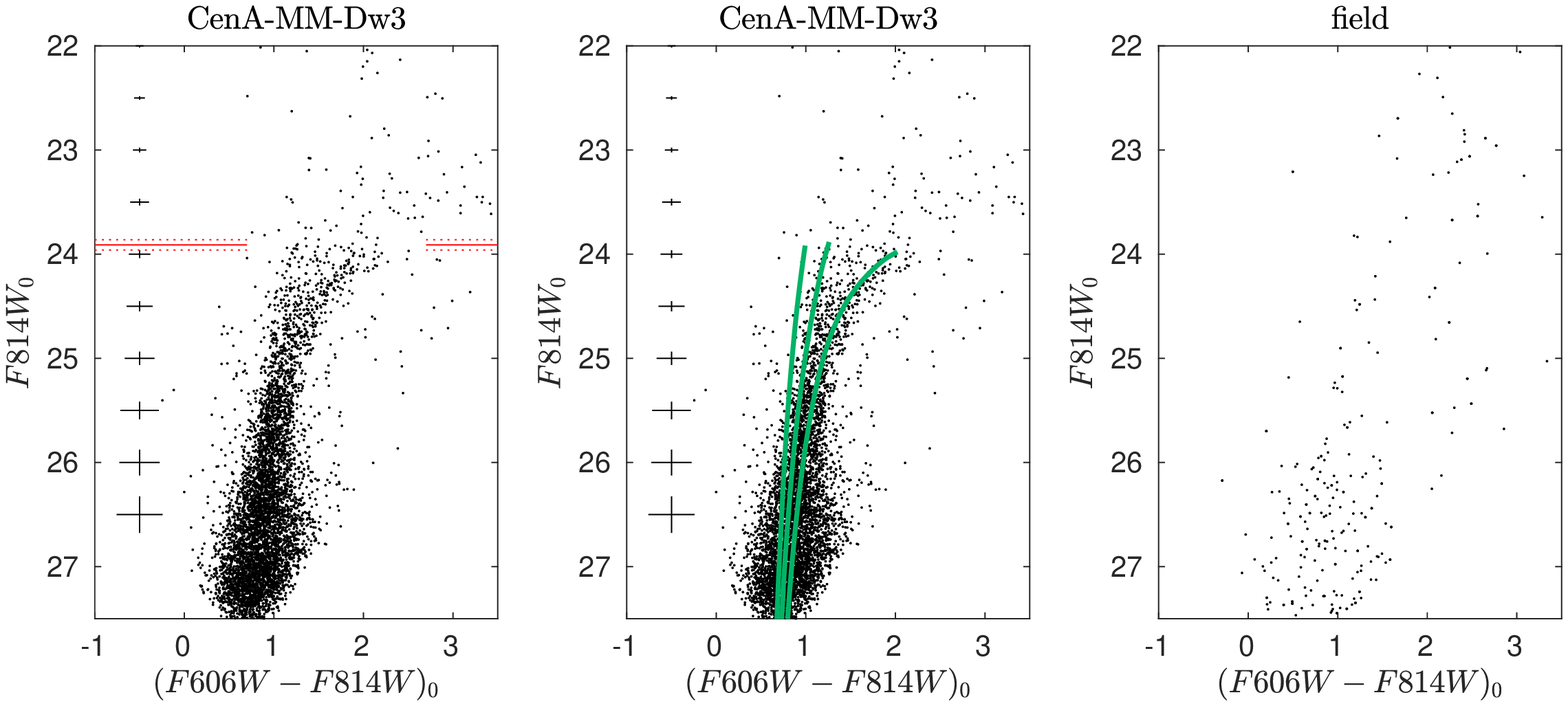}
\caption{Same as Fig.~\ref{dwonetwo}, for the heavily disrupting satellite Dw3. The CMD contains populations within $0.5 r_h$, and the isochrones are for [Fe/H]$=-2.0$, $-1.0$ and $-0.5$.  We discuss this dwarf, and surrounding HST pointings, in some detail in Section~\ref{sec:disrupting} and in Figures~\ref{dw3_map} and \ref{dw3_mdf}.
}
\label{dw3}
\end{figure*}

\begin{figure*}
 \centering
\includegraphics[width=7.cm]{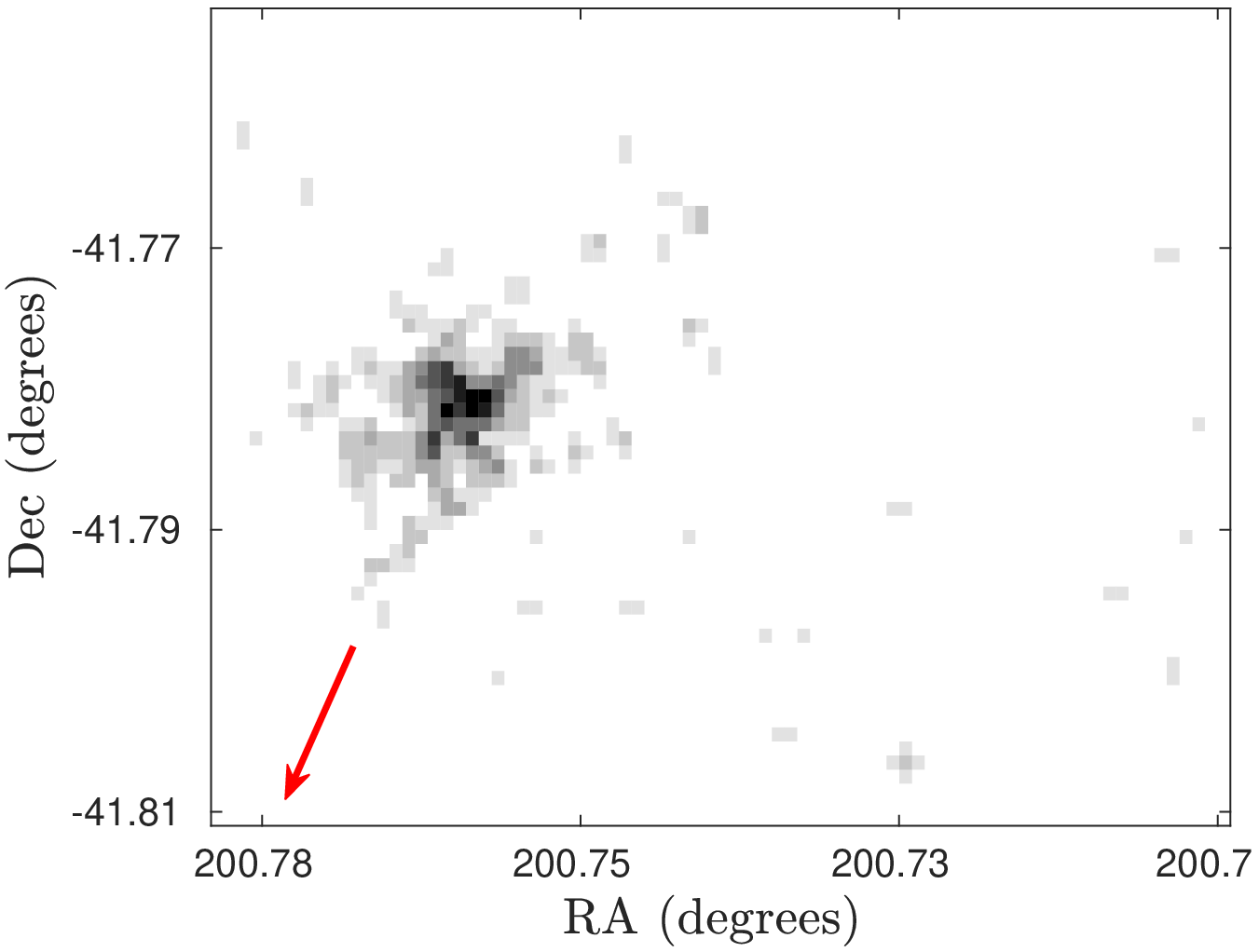}
\includegraphics[width=18.cm]{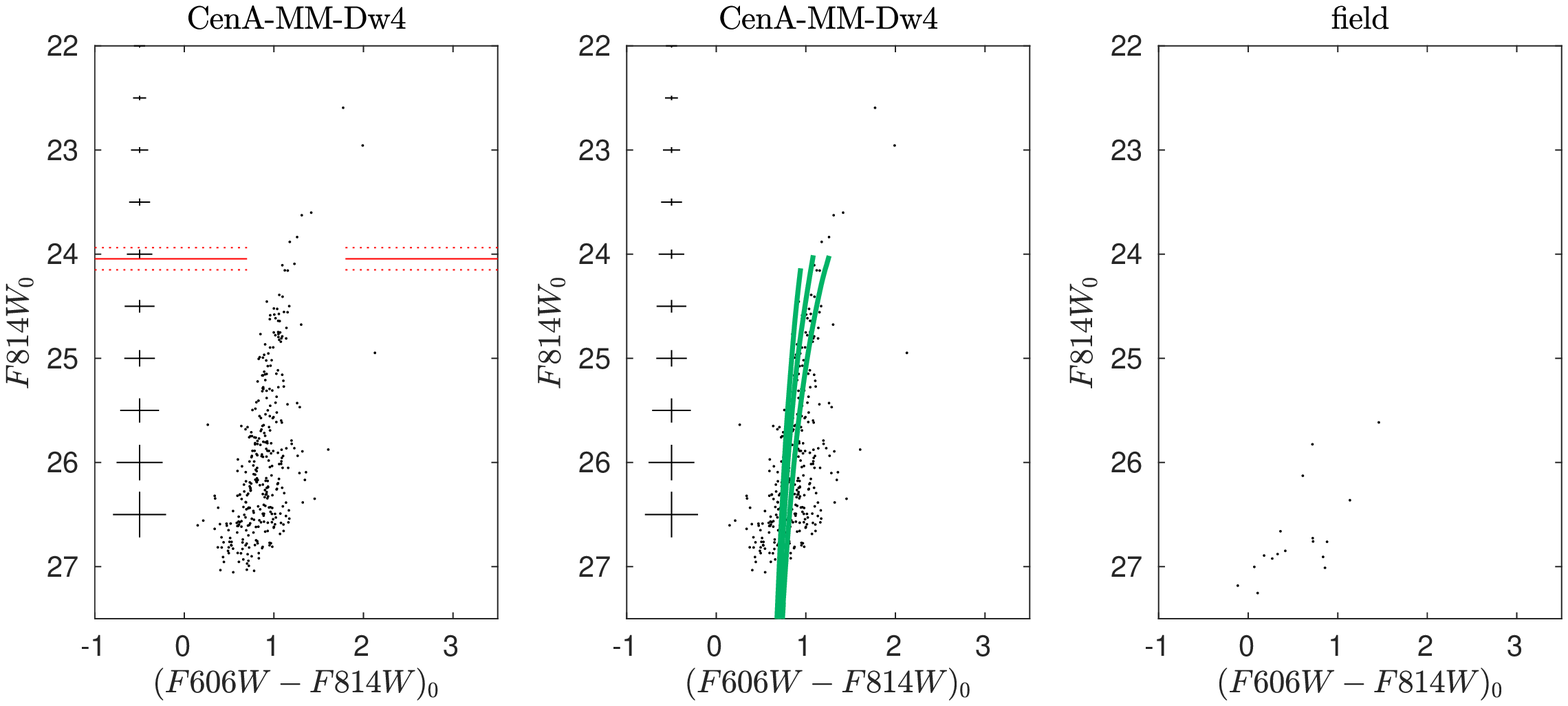}
\caption{Same as Fig.~\ref{dwonetwo}, for Dw4. Stars within $r_h$ are shown in the CMD, and isochrones have metallicities of [Fe/H]$=-2.5$, $-1.5$, and $-1.0$.}
\label{dw4}
\end{figure*}

\begin{figure*}
 \centering
\includegraphics[width=7.cm]{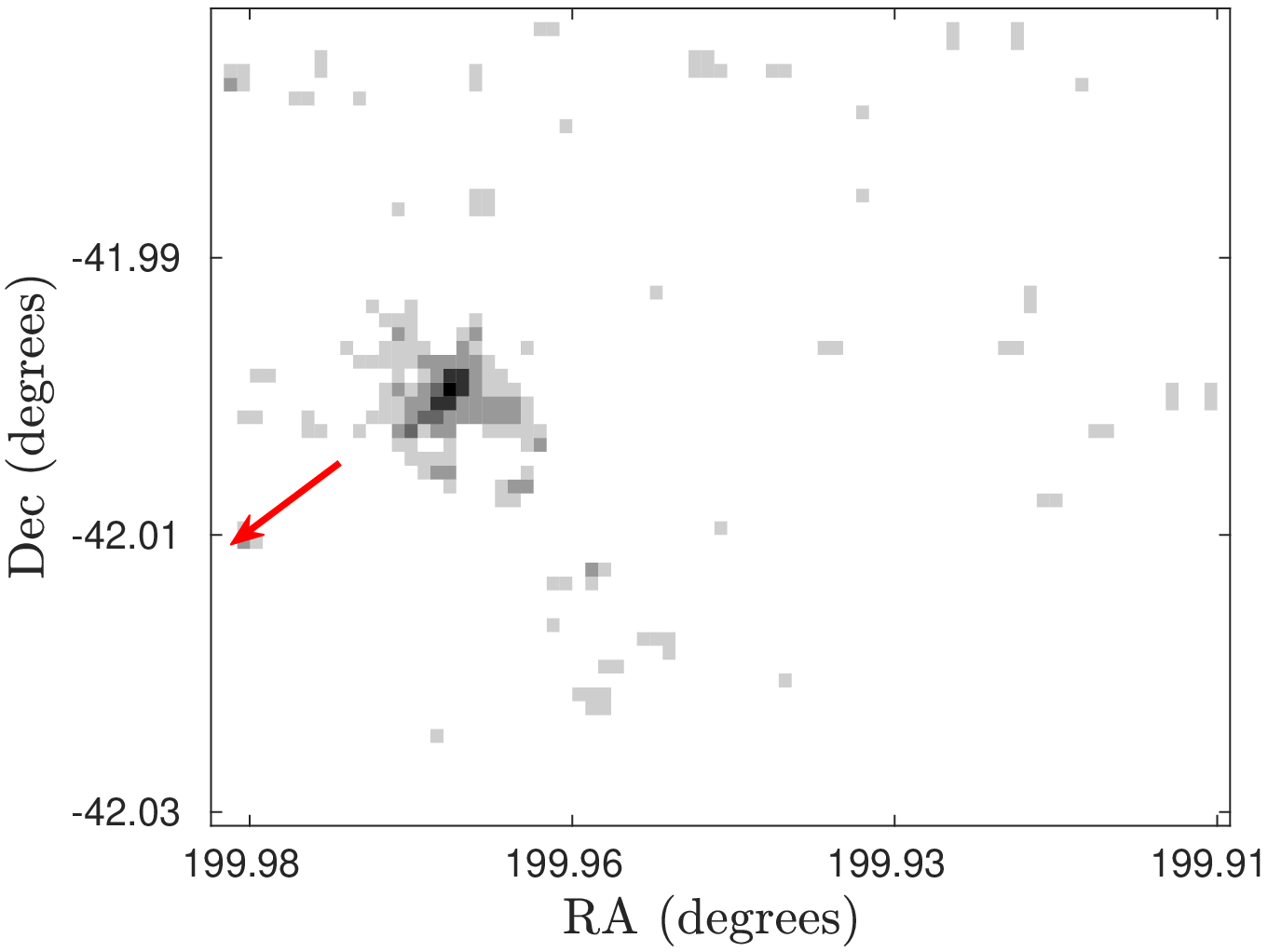}
\includegraphics[width=18.cm]{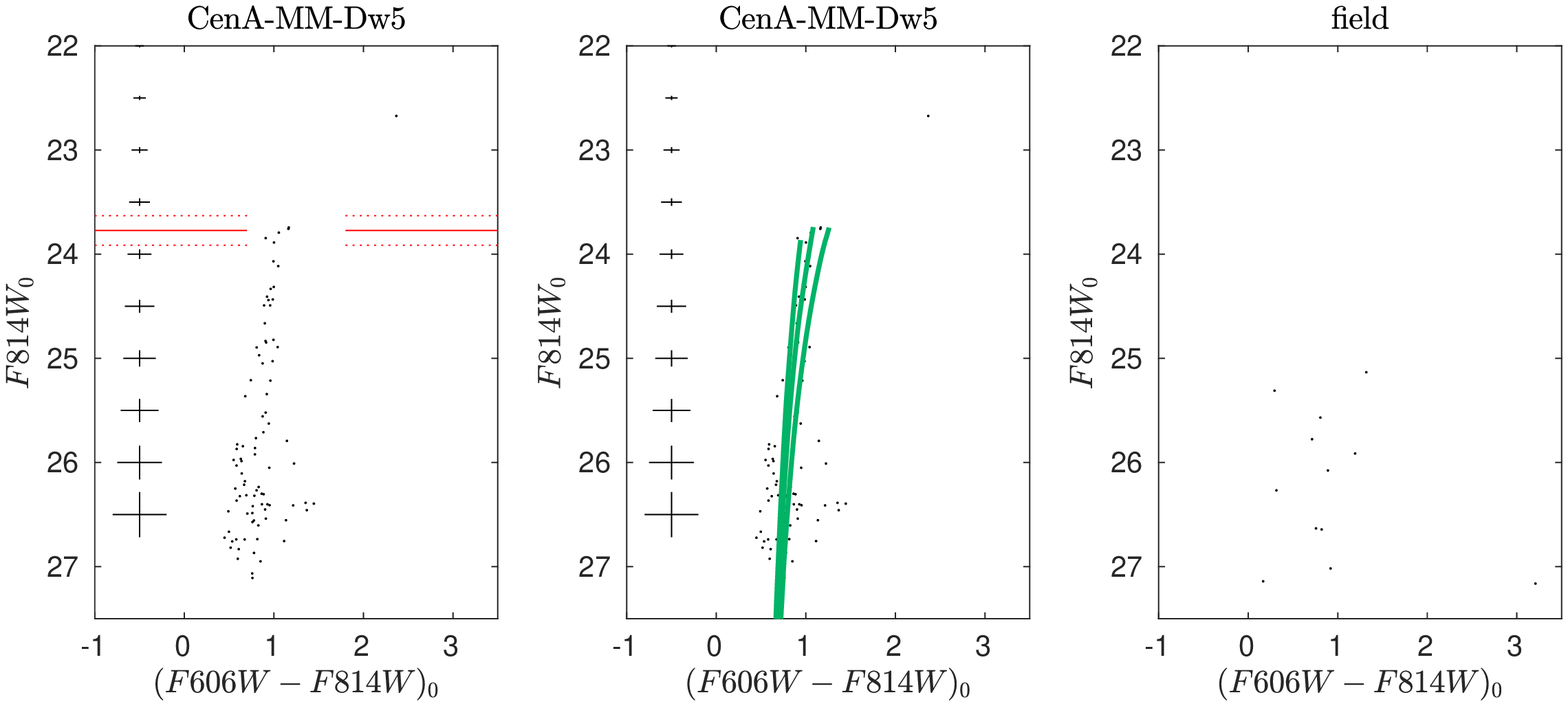}
\caption{Same as Fig.~\ref{dw4}, for Dw5.} 
\label{dw5}
\end{figure*}

\begin{figure*}
 \centering
\includegraphics[width=7.cm]{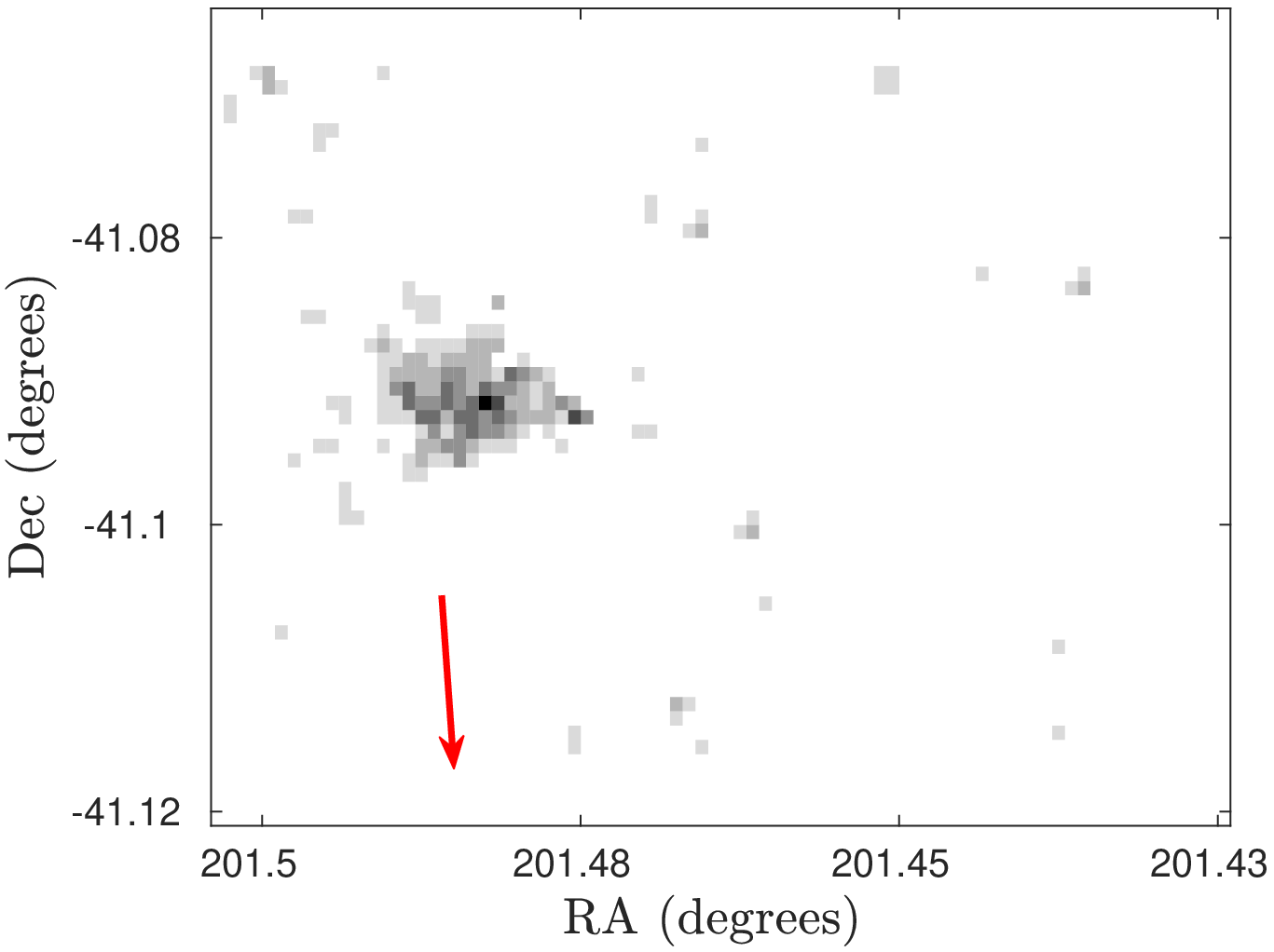}
\includegraphics[width=18.cm]{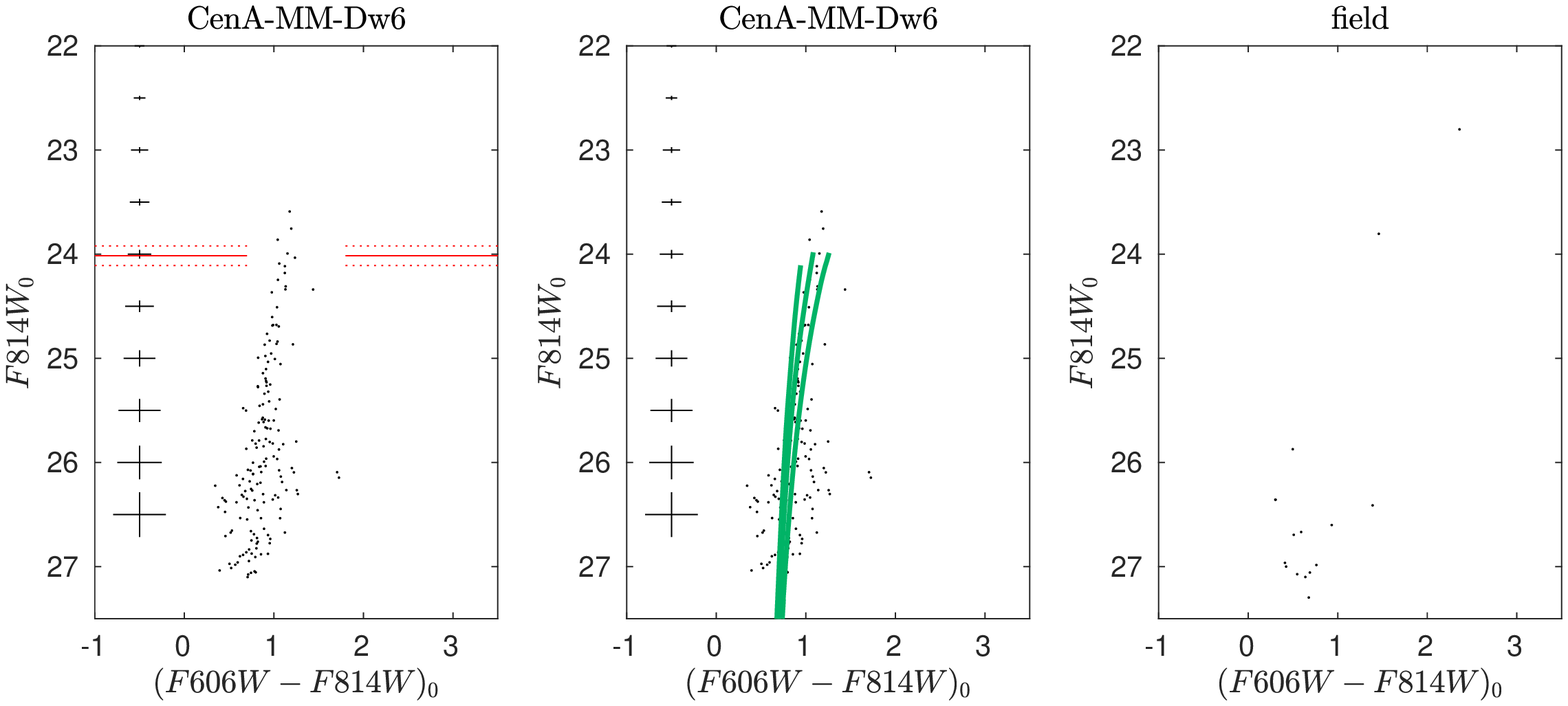}
\caption{Same as Fig.~\ref{dw4}, for Dw6.}
\label{dw6}
\end{figure*}

\begin{figure*}
 \centering
\includegraphics[width=7.cm]{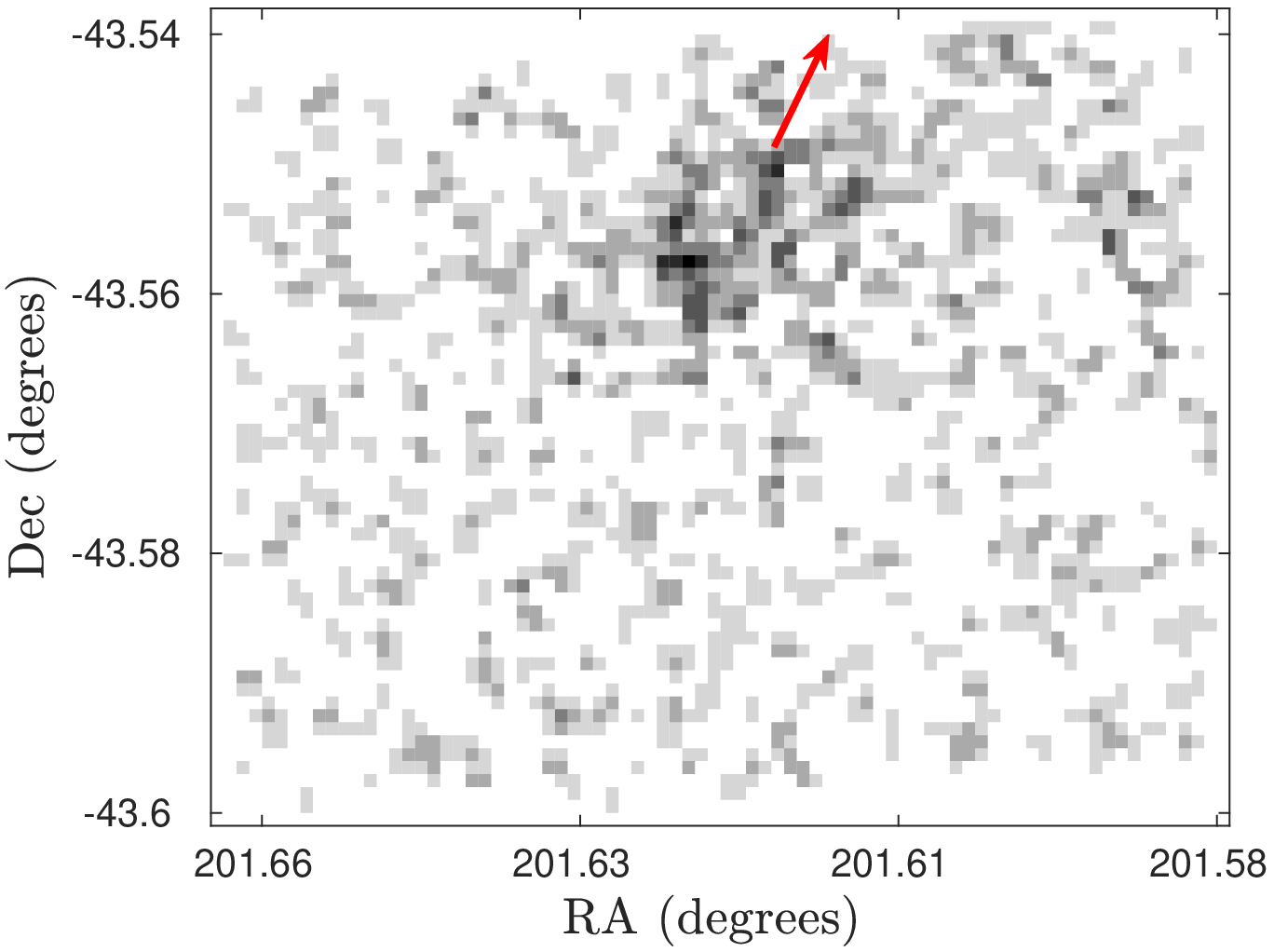}
\includegraphics[width=7.cm]{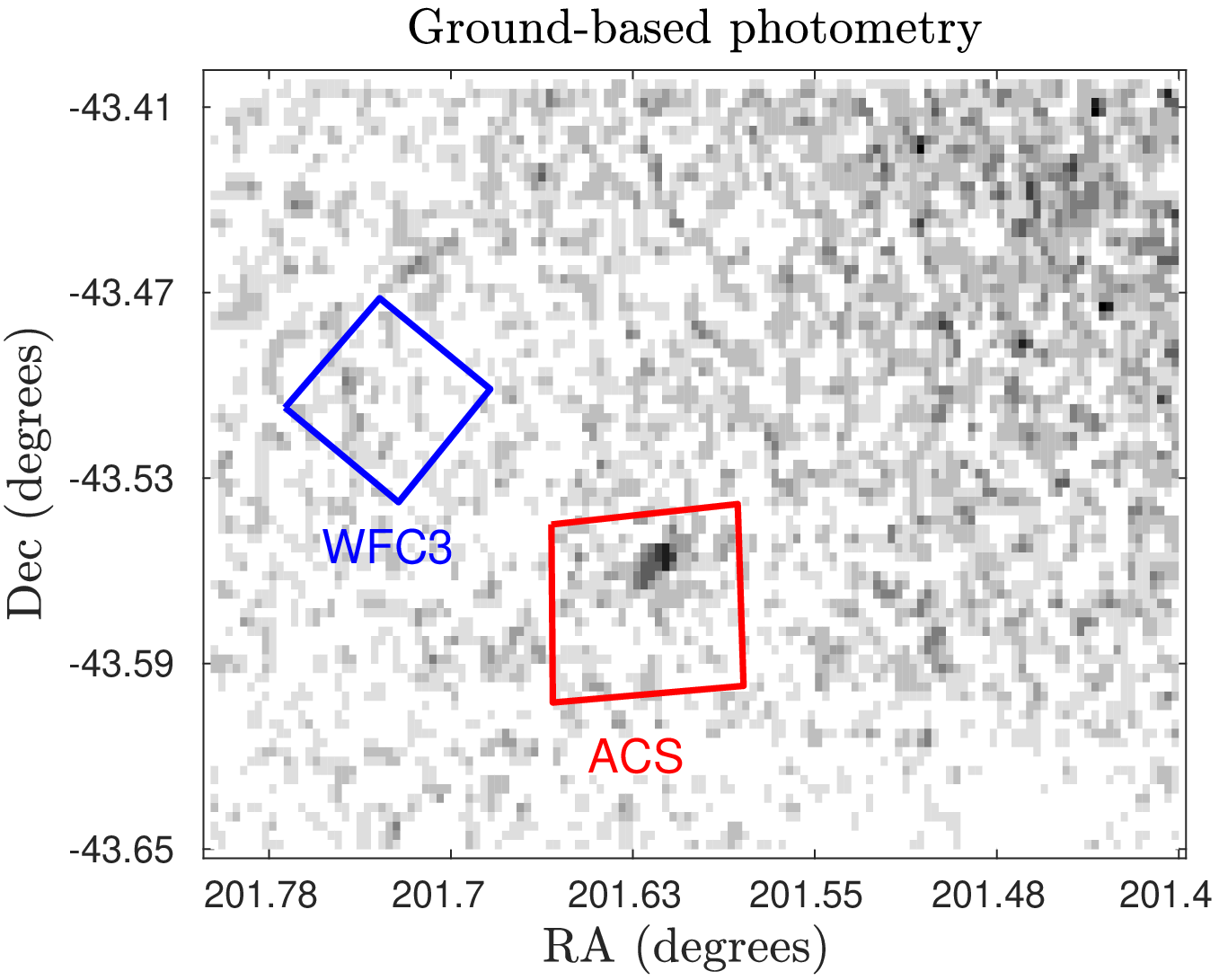}
\includegraphics[width=18.cm]{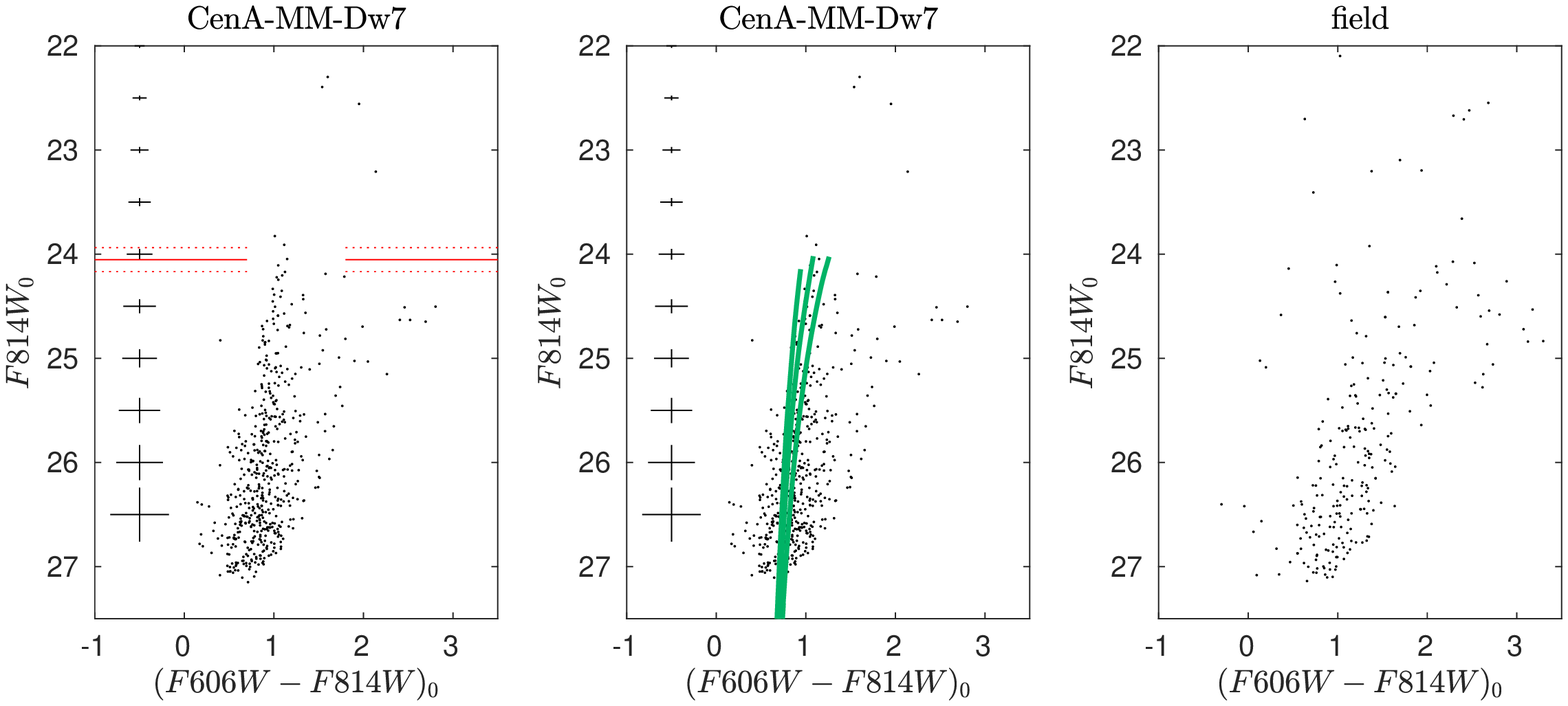}
\caption{Same as Fig.~\ref{dw4}, for Dw7. The additional RGB spatial density map (\emph{upper right panel}) is derived from our ground-based Magellan photometry, and the positions of the ACS and WFC3 pointings are overlaid (see Sect.~\ref{individual}). The contamination from Cen~A's metal-rich stellar populations at the position of Dw7 is clear both from the RGB density map, as well as from the field CMD.}
\label{dw7}
\end{figure*}

Seven Cen~A satellites were resolved into stars in our Magellan imaging \citep{crnojevic14b,crnojevic16}, and have now been followed-up with our HST program and confirmed as group members. The deep CMDs obtained with the ACS camera reveal stellar populations consistent with those expected at the group's distance (see Fig.~\ref{boxes}), and we show the individual dereddened CMDs for dwarfs Dw1 to Dw7 in Figs.~\ref{dwonetwo}--\ref{dw7}.

In \citet{crnojevic16}, we reported on four candidate dwarfs detected
as surface brightness enhancements without a resolved stellar
counterpart (Dw10 to Dw13). Our follow-up HST imaging confirmed the
lack of resolved populations in these targets, thus excluding the
possibility that they are low-mass satellites of Cen~A and are instead
mostly background galaxies. In the case of Dw13, examination of the
HST images (in particular, the lack of a structured surface brightness
enhancement) has led us to conclude that the object is galactic cirrus
and not a genuine background object
\citep[e.g.,][]{guhathakurta89}. We rename these contaminant objects
Dw10-C16 to Dw13-C16 (from the original \citealt{crnojevic16}
nomenclature) to avoid confusion with the new candidate satellites
discovered in our ongoing survey, which are now dubbed Dw10, Dw11, and
so on (see Section~\ref{sec:newdwarfs}). Because of the very low
surface brightness of these background objects, integrated photometry
from the HST images is very challenging, and we thus use the
Magellan/Megacam images to derive their luminosities and structural
properties. We model them with GALFIT \citep{peng02_galfit}, following
the procedure adopted in \citet{bennet17}, and the derived values are
reported in Table~3. Among these unresolved objects, Dw11-C16 is
noteworthy: given its low surface brightness and large effective
radius, this galaxy would be classified as an ultra-diffuse galaxy if
located distances $\gtrsim 26$~Mpc (distances cannot be constrained
for unresolved objects with our dataset); Dw10-C16 would also qualify
as an ultra-diffuse galaxy, however only at larger distances $\gtrsim 80$~Mpc. Finally, we have searched a region around each of these objects in NED to investigate a possible association with known galaxies, but have found them to be isolated.

Given that all unresolved candidate dwarfs identified in PISCeS turned out to be background objects, we conclude that it is unlikely that unresolved candidates in our ground-based imaging are real satellites; this also implies that we expect galaxies with $M_V\lesssim-8.0$ (our faintest detected satellite) to be resolved into stars in our survey.

\subsection{Distances from $HST$ Imaging}  \label{dist_sec}

The TRGB distance measurement method is widely used for nearby galaxies resolved
into stars \citep[e.g.,][]{lee93, sakai97, makarov06, rizzi07}. It relies
on the fact that the $I$-band LF of old and 
metal-poor RGB stars presents a sharp break at its bright end that
is a robust standard candle, insensitive to metallicity.
The absolute magnitude calibration for the TRGB in the HST filter system has 
been recently revised by \cite{jang17}. Their new value is a factor of two 
more accurate than previous estimates \citep[e.g.,][]{rizzi07}.
\cite{jang17} also determined the TRGB color dependence
based on deep HST images of eight nearby galaxies:

\begin{equation}
\begin{array}{lr}
M^{\rm TRGB}_{F814W}= -4.015(\pm0.056)-0.159(\pm0.01) \\
\times[(F606W-F814W)_0-1.1]^2 +0.047(\pm0.02) \\
\times[(F606W-F814W)_0-1.1]
\end{array}
\end{equation}

We apply this color correction term to our photometry in order to obtain a sharper 
and more easily measured 
 \citep[see, e.g.,][]{madore09, mcquinn16}, and 
apply our TRGB detection algorithm to this corrected photometry. When deriving TRGB values, we consider stars within 1--2 half-light radii (the latter case is for dwarfs containing small numbers of stars), and we only consider stars with colors 
$0.7<(F606W-F814W)_0<1.5$. To find the TRGB value, we adopt the approach by \cite{makarov06}, where
a pre-defined LF is compared to the observed RGB LF. The model LF has
the form of two distinct power laws, 

\begin{equation}
 \psi = \left\{ \begin{array}{lr}
                 10^{a(m-m_{\rm TRGB})+b}, & m - m_{\rm TRGB} \ge 0, \\
                 10^{c(m-m_{\rm TRGB})},   & m - m_{\rm TRGB} < 0
                \end{array} \right.
\end{equation}

\noindent where $a$ and $c$ are the slopes of the RGB and AGB, respectively, and
$b$ represents the discontinuity at the TRGB magnitude.
The photometric uncertainty, bias and incompleteness function derived from
the artificial star tests are modeled with continuous functions and
convolved with the pre-defined LF.
We fit the pre-defined function with a non-linear
least squares method, using a Levenberg--Marquardt algorithm. 
We use an initial guess for $a, b$ of 0.3, while for
$m_{\rm TRGB}$ we compute a first estimate with the Sobel filter 
edge-detection technique described in \cite{sakai97}. Briefly,
the observed LF is binned and smoothed with a Gaussian function
following the photometric errors, and then convolved with a 
Sobel filter which highlights the position of the LF edge.
This method depends on the chosen binning of 
the LF, and we thus refine our measurement by fitting the 
model LF to our data. The latter method gives minimal 
differences for a range of bin sizes of the observed LF, and
returns smaller uncertainties on the measured TRGB values.

The TRGB values, the distance moduli and the distances for our targets
are reported in Table~\ref{properties}.
The distances derived from the HST dataset are  slightly higher on average than 
those derived from the discovery Magellan dataset, but mostly consistent within 
the errorbars. The exceptions are: Dw7, which is $\sim0.5$~mag more distant, likely because in the ground-based dataset we could not easily separate Cen~A's strong contamination from the dwarf's population (this is the closest dwarf in projection to Cen~A); and Dw3, which is $\sim0.4$~mag closer likely because the photometric confusion in the Magellan dataset led us to misidentify the brightest RGB stars as luminous AGB stars (the HST CMD shows a less prominent intermediate-age AGB population).
The accuracy of the updated HST distance values is improved by a factor of $2-3$ with respect to the Magellan values reported in \citet{crnojevic16}.

\subsection{Structural parameters and luminosities}  \label{struc_sec}

We derive the structural parameters for our confirmed Cen~A
satellites with the maximum likelihood method presented in
\cite{martin08}, using the implementation in \cite{sand12}.
The code fits the selected RGB stars from the HST data (see Fig.~\ref{boxes})
with an exponential profile with the following free parameters:
central coordinates, position angle, ellipticity, half-light
radius, and background surface density. 
The exponential profile is a good fit for
all of the confirmed dwarfs. We report the resulting parameters in Tab.~\ref{properties};
the uncertainties come from boostrap resampling of the data.

In two cases the adopted procedure for the derivation of structural parameters deviates from that just described. For Dw1, the HST data cannot be used to constrain structural parameters, since this dwarf extends well beyond the ACS FoV, and thus the original Magellan photometry is adopted instead. Moreover, Dw1 hosts a nuclear star cluster, which provides an additional cuspy component to its surface brightness profile (Seth et al., in prep.). For our structural parameter derivation, a small region around the central cluster is thus masked (even though it is not resolved in the ground-based images). For Dw3, which similarly extends well beyond the ACS FoV, and is the central region of a $\sim$60 kpc long disrupting dwarf and tidal stream system, we decided to keep the half-light radius from the Magellan photometry \citep[see][]{crnojevic16}, which was derived from a fit to the remnant galaxy core, i.e., excluding the tidal tail regions. The central coordinates were derived from the deeper HST photometry with an iterative process, computing the average of the stellar positions
within circles of decreasing radius, while ignoring the sources found within a radius of 0.15~arcmin from the central cluster center.  We discuss this disrupting dwarf further in Section~\ref{sec:disrupting}.

\begin{figure*}
 \centering
\includegraphics[width=8.5cm]{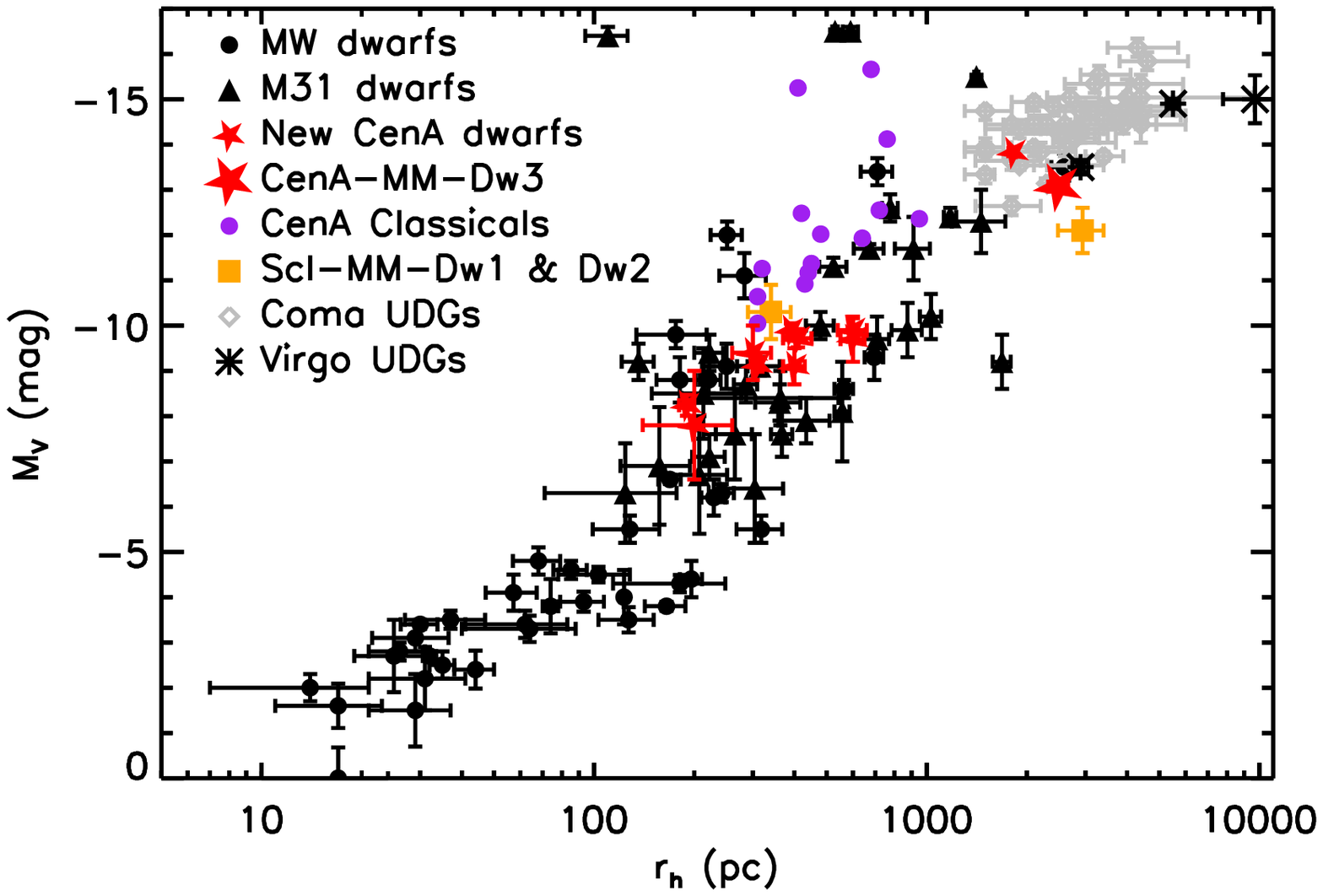}
\includegraphics[width=8.5cm]{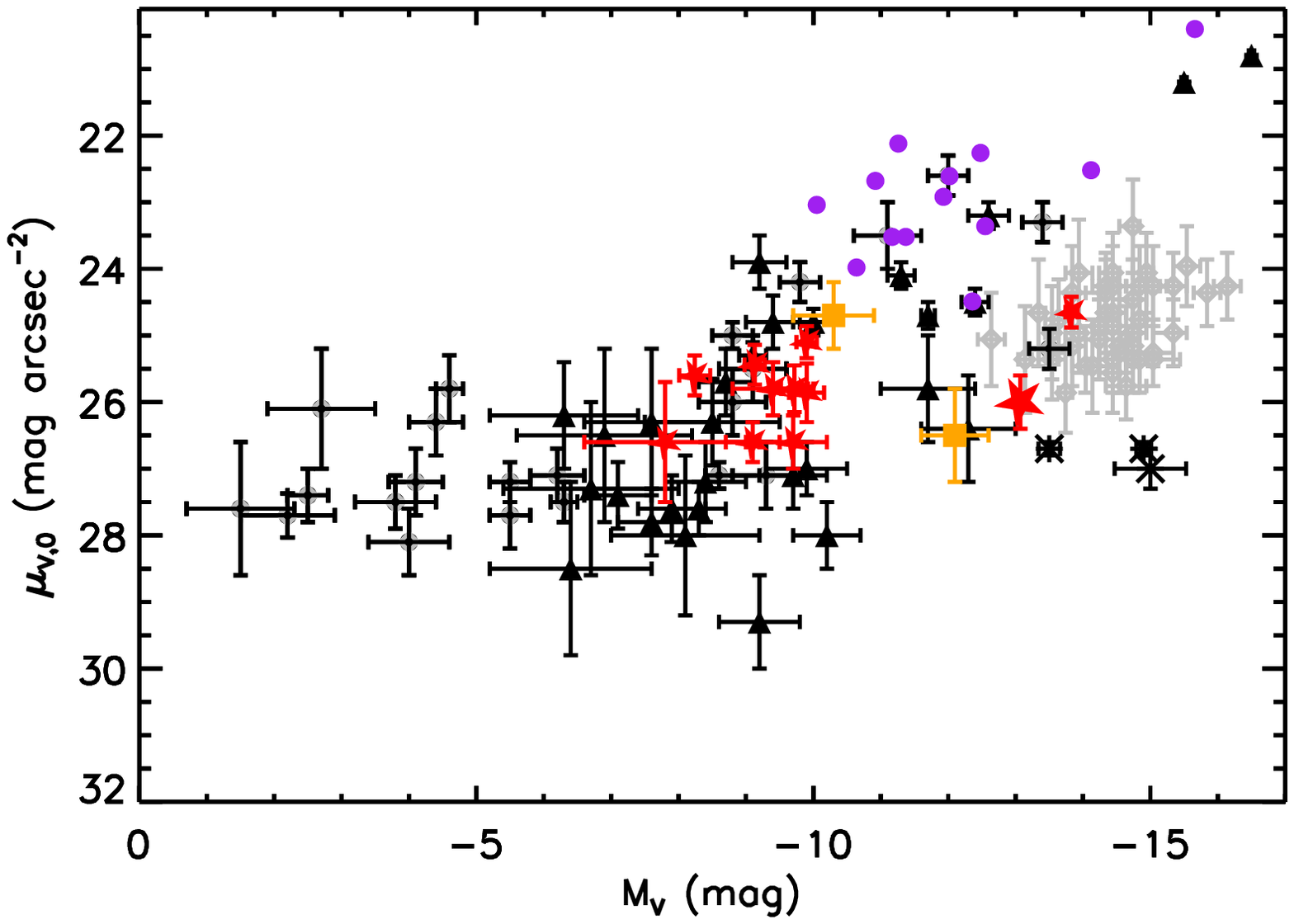}
\caption{Properties of the new Centaurus A satellites with respect to Local Group and other galaxy samples.  \emph{Left panel.} Absolute $V$-band magnitude as a function of half-light radius.  \emph{Right panel.} Central surface brightness as a function of absolute magnitude.  The Cen~A dwarfs presented in this work, with updated physical properties, are shown as red stars. The tidally disrupting CenA-MM-Dw3 is denoted with a larger red star symbol. The general properties of the faint Cen~A dwarfs presented here are consistent with analogous galaxies in the Local Group. They are also fainter and have a lower surface brightness than the previously know Cen~A sample. CenA-MM-Dw1 and the disrupting dwarf CenA-MM-Dw3 are comparable to ultra-diffuse galaxies seen in the Virgo and Coma clusters.  The data for both panels comes from: MW and M31 dwarf galaxies (black points and triangles, respectively; data from \citealt{mcconnachie12,sand12,crnojevic14a,drlica15,kim15a,kim15b,koposov15,laevens15a,laevens15b,martin15,crnojevic16,Drlica16,Torrealba16,Carlin17,Mutlu18,Koposov18,Torrealba18}), recently discovered PISCeS dwarfs in NGC~253 \citep[orange squares;][]{Sand14,toloba16}, diffuse galaxies in Virgo and Coma \citep[black asterisks and gray diamonds, respectively;][]{mihos15,vandokkum15} and previously known ``classical'' Cen~A dwarfs \citep[purple circles;][]{sharina08}.
}
\label{scaling}
\end{figure*}

The absolute magnitude for the confirmed dwarfs is derived as follows. We produce a
well-populated ($5\times10^6$ to $5\times10^8 M_\odot$, depending on the dwarf) fake population in the HST filters by interpolating
Padova isochrones with an old age (10~Gyr) and with the median
metallicity derived below for each of the galaxies (see Section~\ref{mets_sec}), assuming
a Kroupa IMF \citep{kroupa01}. The fake populations are convolved
with photometric errors as derived from the artificial star
tests. We then extract stars randomly from
this fake population by rescaling the number of stars in the RGB
selection box to the observed number of RGB stars within the half-light
radius (after subtracting field contaminants). The flux of the 
extracted fake stars is summed up along the entire LF, in order to account for the faint, unresolved 
component of the galaxy, and the total luminosity is obtained
by multiplying this quantity by two. This process is
repeated 500/1000 times in order to assess uncertainties;
the absolute magnitudes, derived using our computed
distance moduli, are converted to a Vegamag $V$-band
value by following the \cite{sirianni05} prescriptions.
We further calculate the central surface brightness values starting
from the derived absolute magnitude and the half-light radius,
assuming an exponential profile. The final values can be found in Table~\ref{properties}, and Fig.~\ref{scaling} shows the relation between the derived absolute magnitudes, half-light radii and central surface brightnesses as compared to Local Group dwarfs and other galaxy samples.

Overall, the structural parameters and the luminosities derived from the HST
dataset agree well with the results from the ground-based Magellan 
imaging to within the uncertainties. For Dw3, Dw4, and Dw6, the absolute magnitudes agree to $\sim0.1$~mag. For Dw5 and Dw7, the HST values are about 1~mag brighter (after factoring in the increased distance for Dw7, as well), but the Magellan estimates had large uncertainties at the outset. Dw1 is an extreme case ($M_V\sim-11$ from ground-based photometry and a revised value of $M_V\sim-14$ from HST): we discovered that, at the position of this dwarf, the automated Megacam pipeline provided a significant overestimation in the sky value due to the large size of Dw1. The sky subtraction in the resulting stacked images was thus excessive and led us to measure a lower total luminosity and surface brightness (the central surface brightness was inferred from integrated light, and the absolute luminosity was derived starting from the central surface brightness value and an exponential profile), which we revise with the HST dataset by using resolved stars rather than integrated light. We note that Dw1's structural parameters and luminosity are now in agreement with the analysis of \citet{mueller17} as well. Dw2 (at only 3~arcmin from Dw1) was similarly affected, although to a lesser extent.

\tabletypesize{\scriptsize}
\begin{deluxetable*} {@{}l@{}c@{}c@{}c@{}c@{}c@{}c@{}c@{}}
\tablecolumns{8}
\tablecaption{Properties of the confirmed Cen~A satellites.}

 \tablehead{\colhead{Parameter}  & \colhead{Dw1} &\colhead{Dw2} &\colhead{Dw3} &\colhead{Dw4} &\colhead{Dw5} &\colhead{Dw6} &\colhead{Dw7}}

\startdata
RA (h:m:s) & 13:30:14.31$\pm1.52$''\tablenotemark{*}  & 13:29:57.42$\pm0.98$'' & 13:30:20.44$\pm1.00$''& 13:23:02.56$\pm0.55$''& 13:19:52.42$\pm0.79$''& 13:25:57.25$\pm1.15$''& 13:26:28.55$\pm1.82$''\\
Dec (d:m:s) & $-41$:53:34.78$\pm1.46$''\tablenotemark{*}  & $-41$:52:23.70$\pm1.08$'' & $-42$:11:30.27$\pm11.00$'' & $-41$:47:08.95$\pm0.61$''& $-41$:59:40.68$\pm0.65$''& $-41$:05:37.13$\pm0.80$''& $-43$:33:23.07$\pm1.78$''\\
$E_{(B-V)}$ & $0.11$ & $0.11$ & $0.10$ & $0.11$ & $0.11$ & $0.10$ & $0.10$ \\
$(m-M)_0$ (mag) & $27.96\pm0.07$ & $28.09\pm0.12$ & $27.94\pm0.09$ & $28.06\pm0.14$ & $27.79\pm0.19$ & $28.03\pm0.11$ & $28.07\pm0.15$ \\
D (Mpc) & $3.91\pm0.12$& $4.14^{+0.24}_{-0.22}$& $3.88^{+0.16}_{-0.15}$ & $4.09^{+0.26}_{-0.25}$&  $3.61^{+0.34}_{-0.31}$& $4.04^{+0.20}_{-0.19}$ &$4.11^{+0.29}_{-0.27}$\\
D$_{\rm CenA, proj}$ (deg) & $1.43$ & $1.42$ & $1.22$ & $1.31$& $1.45$& $1.93$& $0.57$\\
D$_{\rm CenA, proj}$ (kpc) & $93$ & $92$ & $79$ & $85$& $94$& $125$& $37$\\
$\epsilon$ & $0.22\pm0.02$\tablenotemark{a}& $<$0.17 & $0.29\pm0.19$\tablenotemark{*,a}  & $0.32\pm0.05$ & $<$0.20 & $0.25\pm0.08$ & $0.41\pm0.08$ \\
P.A. (N to E; $^o$) & $51.1\pm6.1$\tablenotemark{*}& $-$ & $-$  & $-36.8\pm4.3$ & $-$ & $86.9\pm9.5$ & $-46.1\pm6.5$ \\
$r_{h}$ (arcmin) & $1.60\pm0.03$\tablenotemark{*} & $0.34\pm0.03$ & $2.21\pm0.15$\tablenotemark{*,a}  & $0.33\pm0.01$ & $0.18\pm0.01$ & $0.26\pm0.01$& $0.50\pm0.05$ \\ 
$r_{h}$ (kpc) & $1.82\pm0.03$& $0.41\pm0.04$ & $2.49\pm0.17$ & $0.39\pm0.01$ & $0.19\pm0.01$ & $0.31\pm0.01$& $0.60\pm0.06$ \\ 
$\mu_{V,0}$ (mag~arcsec$^{-2}$) & $24.7\pm0.2$ & $25.8\pm0.4$ & $26.0\pm0.4$ & $25.1\pm0.2$ &$25.6\pm0.3$ & $25.4\pm0.3$& $25.9\pm0.4$ \\ 
$M_V$ (mag) &$-13.8\pm0.1$ &$-9.7\pm0.2$ &$-13.1\pm0.1$\tablenotemark{b} &$-9.9\pm0.2$& $-8.2\pm0.2$& $-9.1\pm0.2$&$-9.9\pm0.3$ \\
$L_*$ ($10^5 L_\odot$) &$283.1\pm26.1$ &$6.5\pm1.2$ &$148.6\pm13.7$ &$7.8\pm1.4$ &$1.6\pm0.3$&$3.7\pm0.7$&$7.8\pm2.2$\\
$M_{HI}$\tablenotemark{c} ($10^6 M_\odot$) & $\lesssim5.5$ & $\lesssim6.2$ & $\lesssim4.3$ & $\lesssim5.1$ & $\lesssim3.8$ & $\lesssim4.6$ & $\lesssim6.8$\\
$M_{HI}/L_*$ ($M_\odot/L_\odot$) & $\lesssim0.2$ & $\lesssim9.5$ & $\lesssim0.3$ & $\lesssim6.6$ & $\lesssim23.4$ & $\lesssim12.5$ & $\lesssim8.8$\\
$[$Fe$/$H$]_{\rm med}$ & $-1.02\pm0.01$ & $-1.58\pm0.07$  & $-0.61\pm0.01$\tablenotemark{b} & $-1.15\pm0.01$ & $-1.46\pm0.02$ & $-1.20\pm0.01$ & $-1.47\pm0.05$ \\
\enddata

\tablenotetext{*}{These values have been derived starting from our Magellan/Megacam photometry (the HST data do not cover the entire extent of these galaxies).}

\tablenotetext{a}{From \citet{crnojevic16}; this is an indicative value only, since the studied galaxy is being heavily disrupted.}
\tablenotetext{b}{Excluding tidal tails.}
\tablenotetext{c}{5~$\sigma$ upper limits from HIPASS.}
\label{properties}

\end{deluxetable*}

\subsection{Metallicities}  \label{mets_sec}

We estimate the metallicity content of our target dwarfs by assuming that they host predominantly old and coeval populations, which holds true in the absence of significant young and/or intermediate-age populations (e.g., bright main sequence and/or AGB stars). Under this assumption, the primary driver of the RGB's color is its intrinsic metallicity, which can thus be computed to first order with photometric information alone. This method is robust for
predominantly old populations: the difference between mean spectroscopic and mean photometric metallicities is only $\sim0.1$~dex for old Local Group dwarfs \citep[e.g.,][]{lianou11}; differences of up to $\sim0.5$~dex are observed in case of prolonged star formation histories. For each galaxy, we derive photometric metallicities for each RGB star brighter than $F814W_0=25.5$ (where photometric errors are smaller than $\sim0.15$~mag and the isochrones are most separated in color) by interpolating between Dartmouth isochrones \citep{dotter08} with a fixed age of 10~Gyr and solar-scaled ([$\alpha$/Fe]=0) metallicities in the range [Fe/H]$=-2.5$ to $=-0.5$ \citep[for more details, see][]{crnojevic10, crnojevic13}.
While the choice of age is arbitrary, a slightly younger age would not have a major impact on our results (e.g., adopting 8~Gyr isochrones would change the median metallicity values by $\sim5\%$; \citealt{crnojevic10}); moreover, deep photometric studies of Local Group dwarfs confirm that they all contain old populations \citep[e.g.,][]{weisz14}. We take into account the foreground/background contamination for the resulting metallicity distribution functions (MDFs) by subtracting the "MDFs" obtained for stars in the same CMD space (which, in the case of foreground stars, are not necessarily RGB stars) of a spatial region next to each dwarf.

\begin{figure}
 \centering
\includegraphics[width=8.5cm]{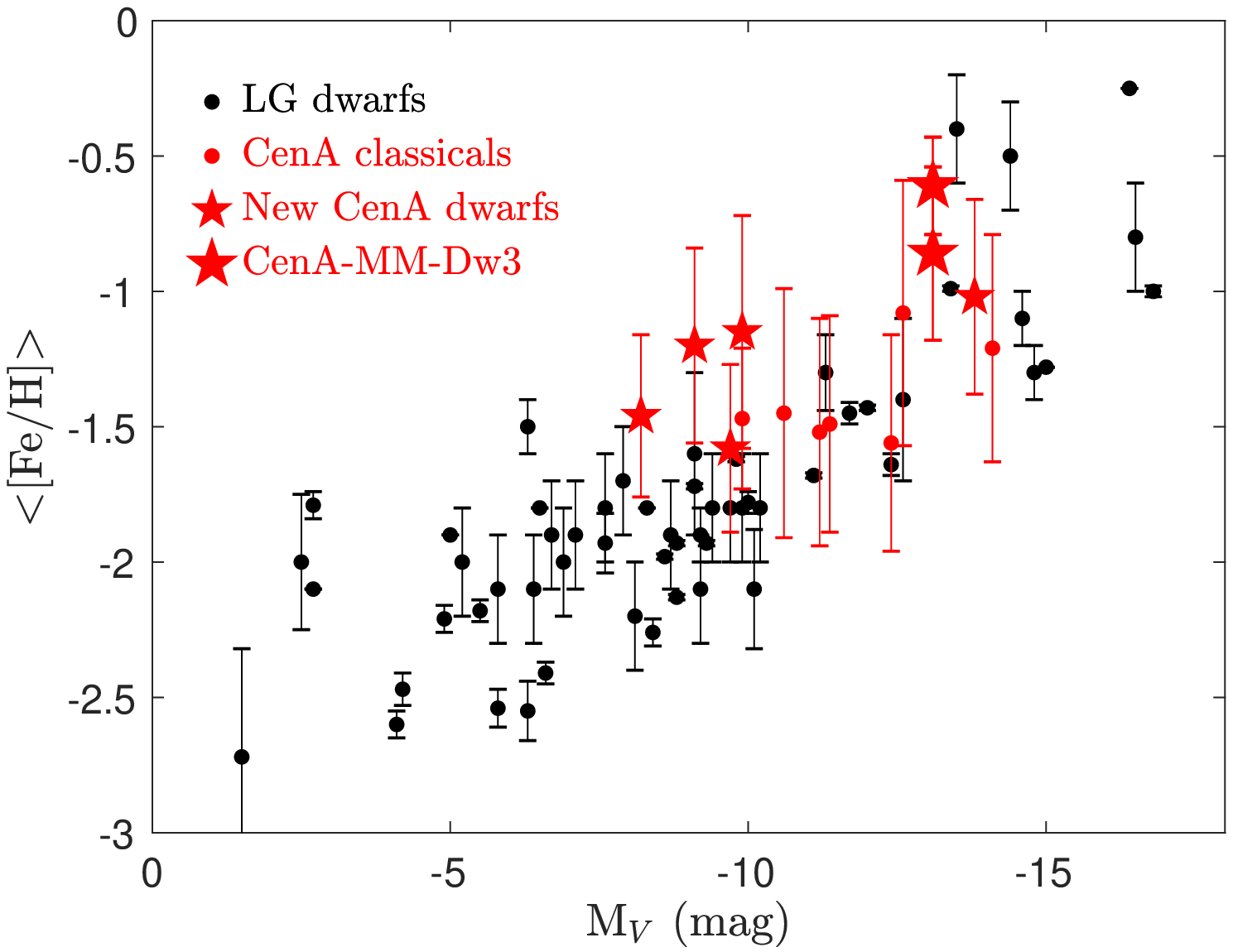}
\caption{Mean/median metallicity versus absolute magnitude for Local Group dwarfs and Cen~A satellites, respectively. The values for Local Group dwarfs are taken directly from \citealt{mcconnachie12}, and are derived with a broad range of techniques. The "classical" Cen~A dwarfs are those studied in \citet{crnojevic10}, for which metallicities are derived with the same photometric procedure as in this study (the other Cen~A satellites do not have reliable metallicity measurements to the best of our knowledge). For Dw3, we report median metallicities for both its remnant and for the most metal-poor pointing along its tail (Dw3S; see Fig.~\ref{dw3_map}). The errorbars for the PISCeS dwarfs' metallicities represent the gaussian spreads of their MDFs.
}
\label{lummet}
\end{figure}

The resulting median metallicities are reported in Table~\ref{properties}. 
For almost all our targets, the MDFs are well approximated by a Gaussian, with the exception of Dw3, which we further discuss below. The Cen~A satellites overall follow the luminosity--metallicity relation defined by Local Group dwarfs \citep[see][]{mcconnachie12}, as shown in Fig.~\ref{lummet}.

\begin{figure*}
 \centering
\includegraphics[width=7cm]{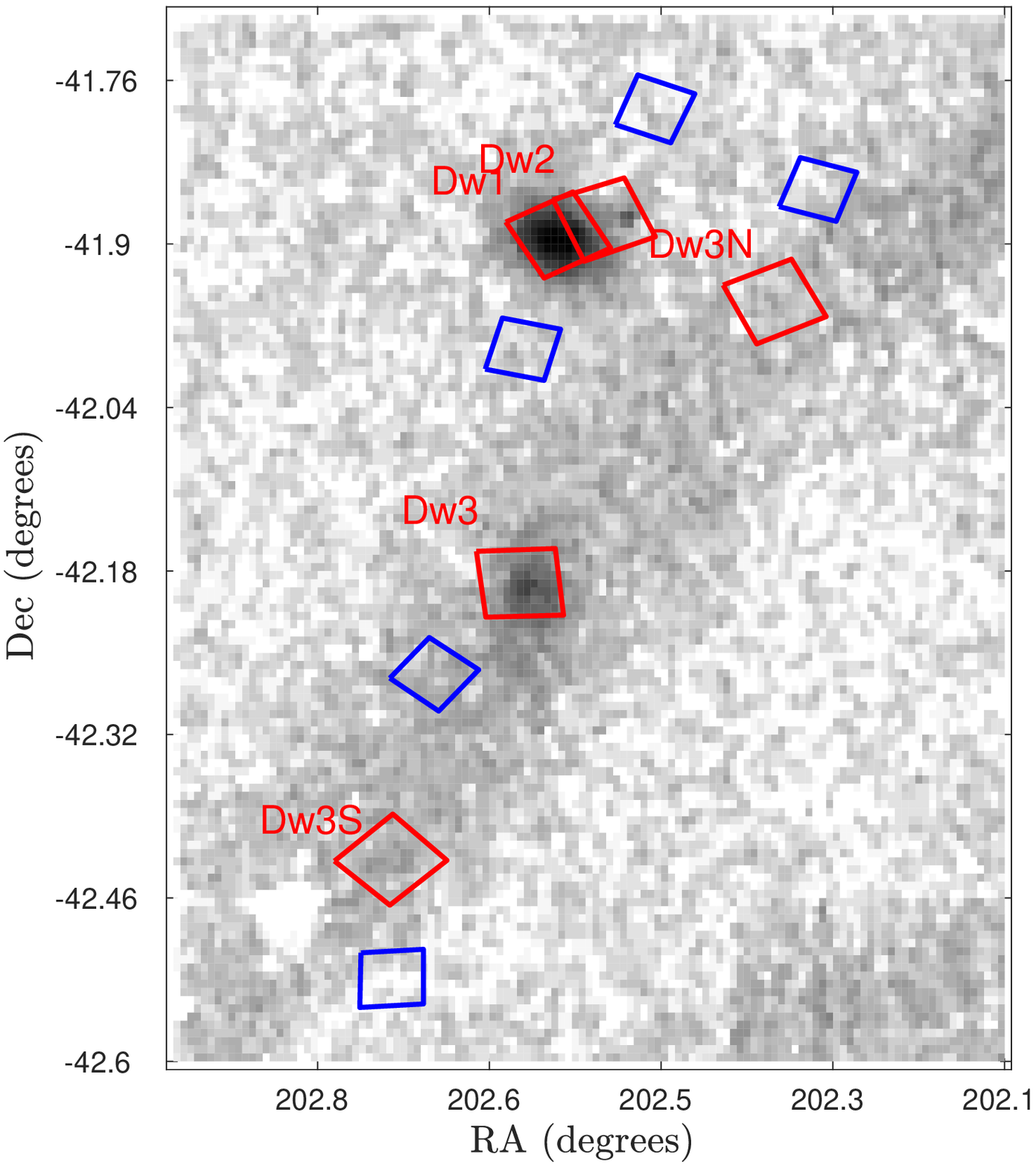}
\includegraphics[width=10.8cm]{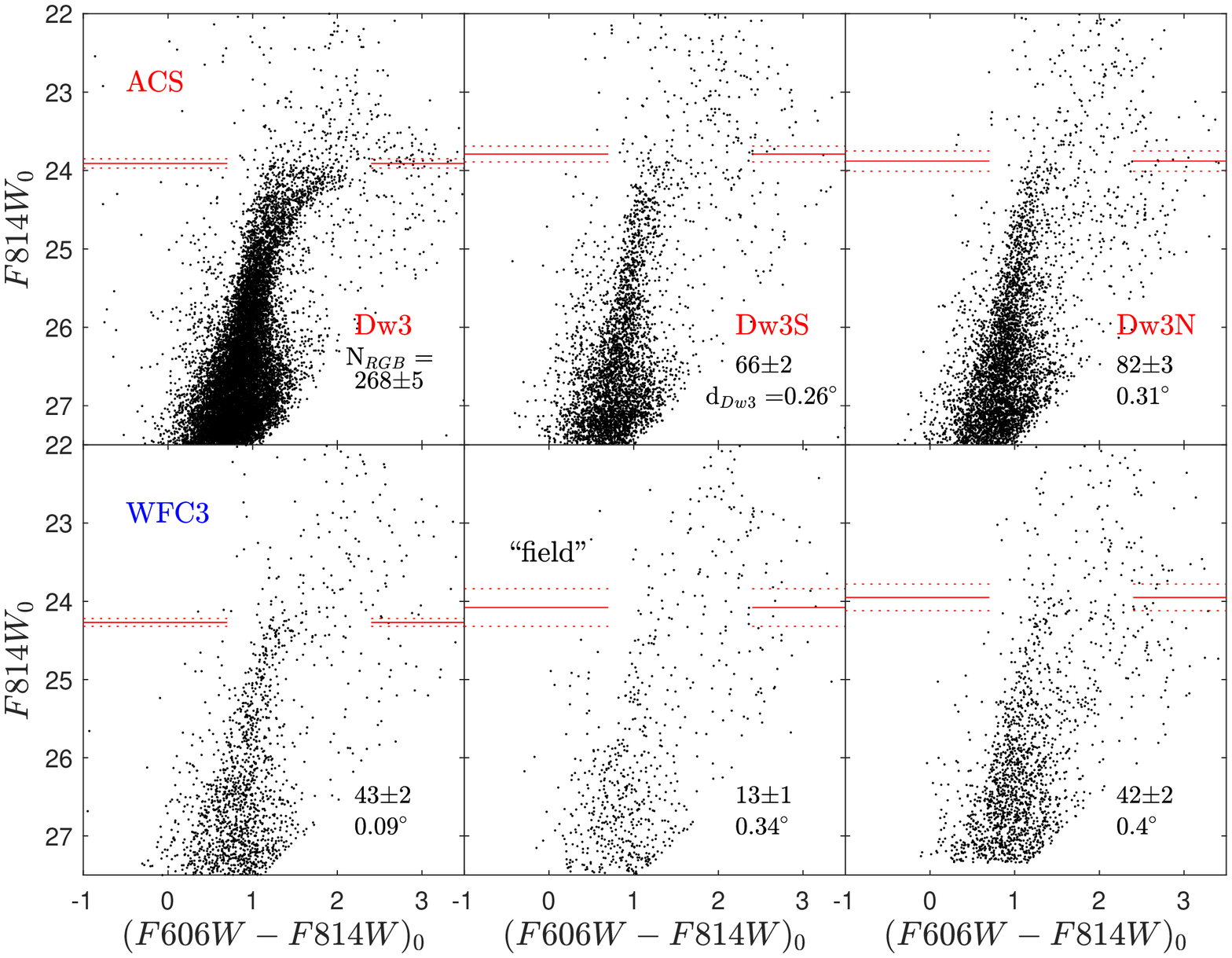}
 \includegraphics[width=10cm]{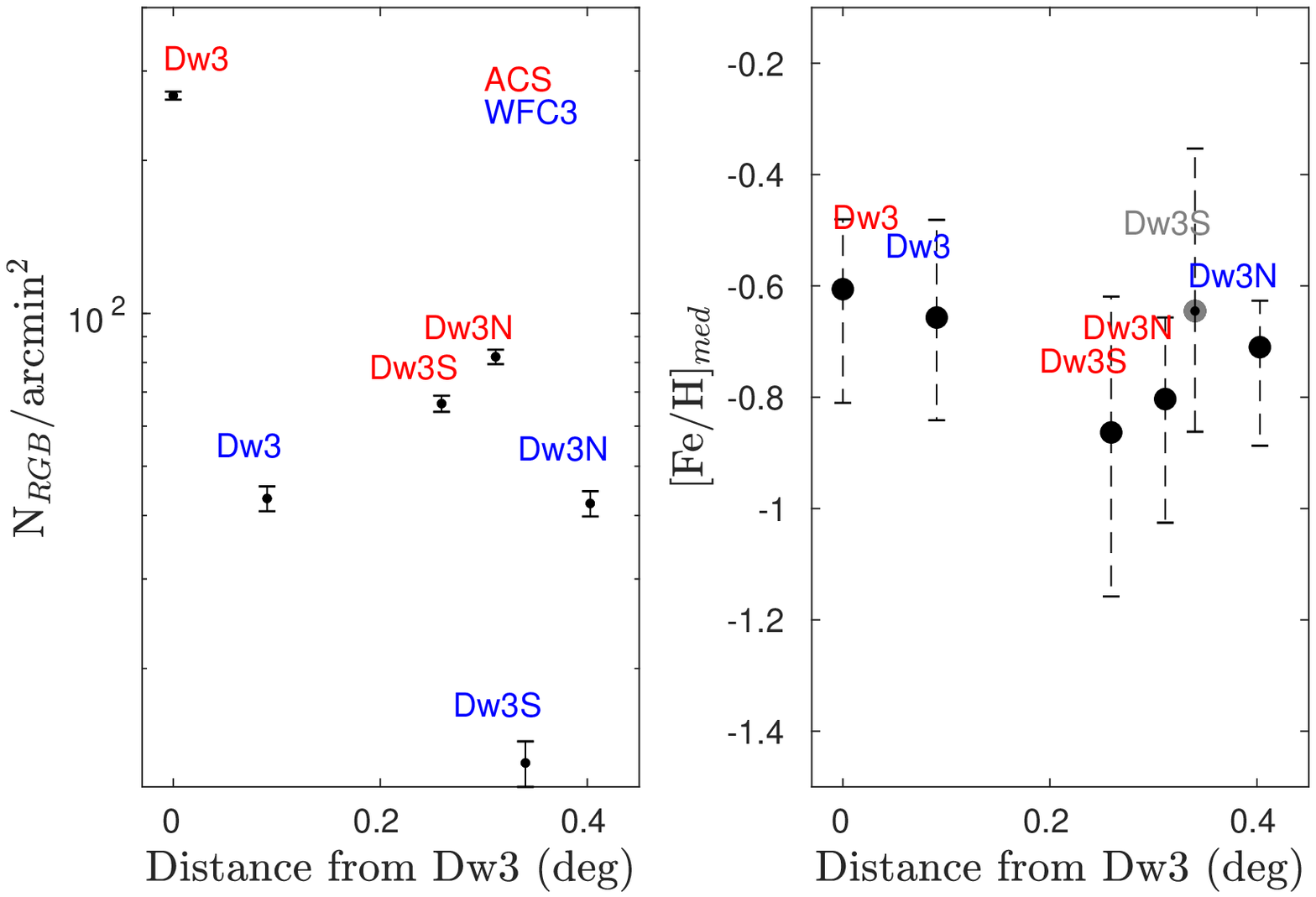}
\includegraphics[width=5cm]{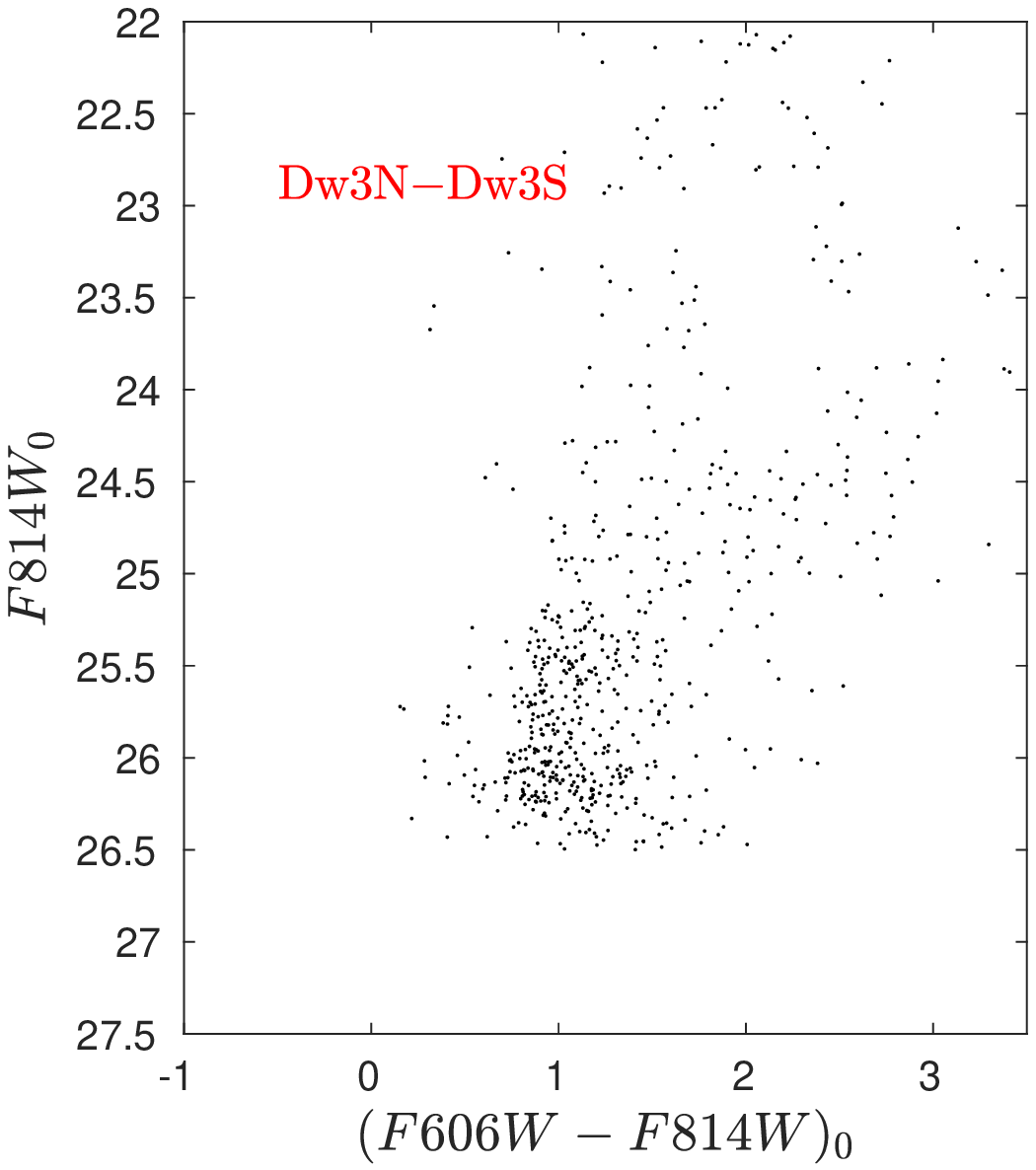}
\caption{\emph{Upper left panel.} Position of the HST fields in the Dw3 area, overlaid on the RGB spatial density map from our ground-based Magellan photometry. Red squares are for the primary ACS/WFC pointings, blue for the parallel WFC3/UVIS ones (the parallel pointing for each primary is the closest blue square to it). \emph{Upper right panels.} CMDs for the HST fields related to Dw3, shown in the upper left panel and labeled accordingly (the primary ACS pointings are in the upper panels, while the corresponding parallel WFC3 pointings are below them). The number of stars per square arcmin in the RGB selection box is reported for each field, as is the projected distance (in deg) of each pointing from the center of Dw3. The TRGB magnitudes are also reported as red lines.
\emph{Lower left and central panels}. Number of RGB stars per square arcmin and median metallicity as a function of distance from the center of Dw3, labeled as above. The metallicity errorbars denote 50th percentile intervals for the respective MDFs. \emph{Lower right panel.} CMD obtained after statistical subtraction of Dw3S pointing stars from Dw3N pointing.
}
\label{dw3_map}
\end{figure*}

\begin{figure*}
 \centering
\includegraphics[width=10.8cm]{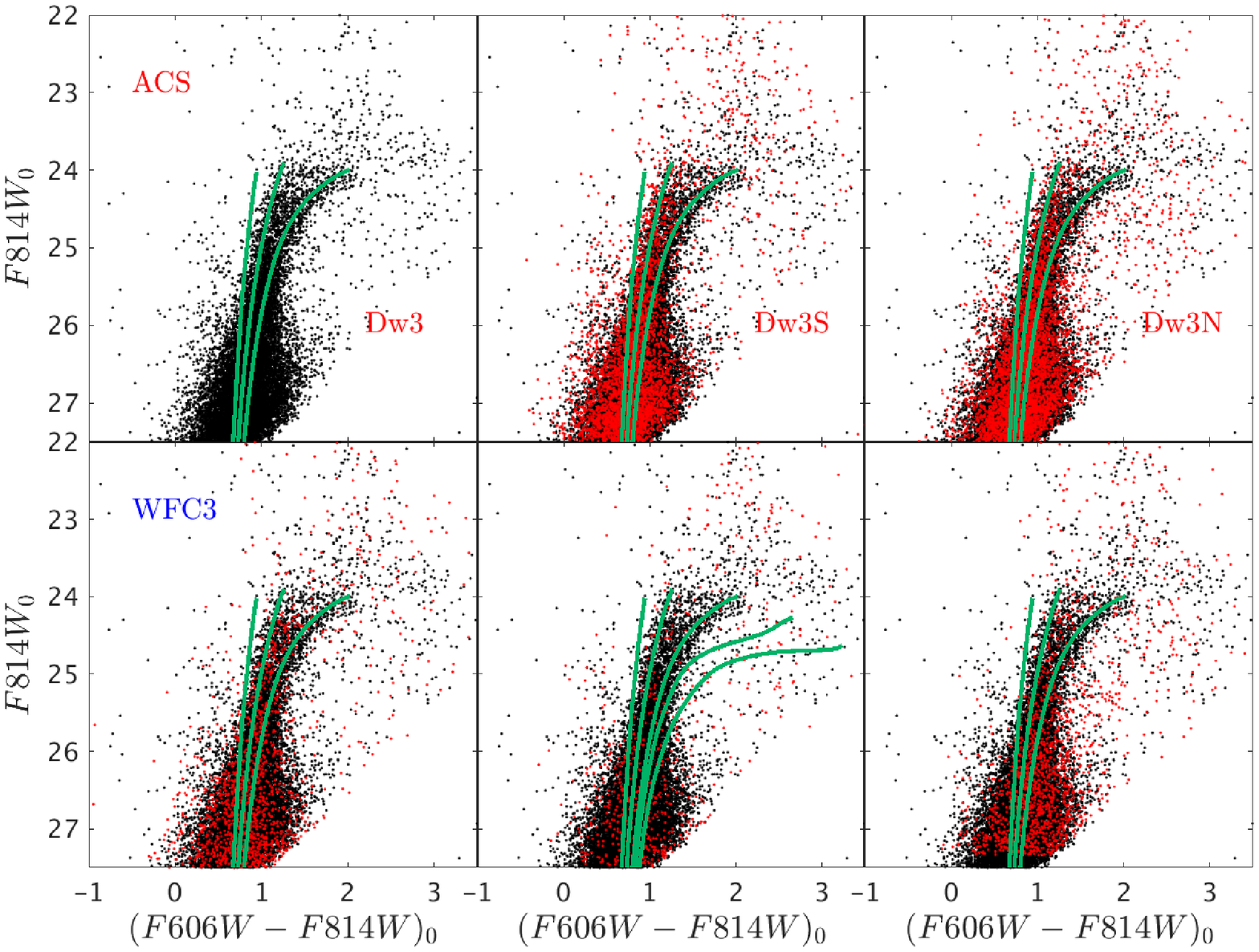}
\includegraphics[width=16cm]{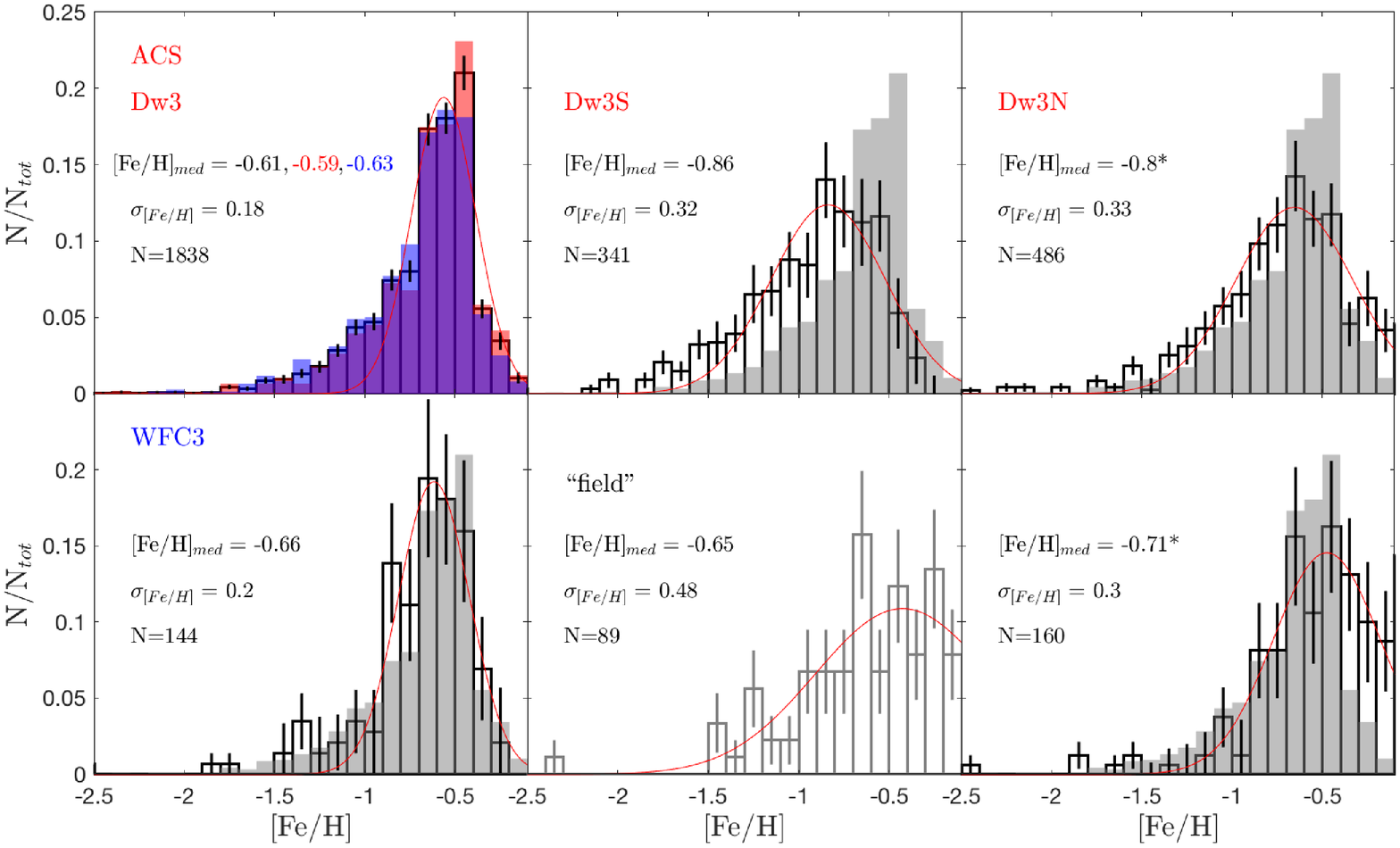}
\caption{\emph{Upper panels.} Same CMDs as in Fig.~\ref{dw3_map}, with overlaid 10~Gyr isochrones of varying metallicity ([Fe/H]$=-2.0$, $-1.0$ and $-0.5$; plus $-0.3$ and $-0.1$ for the central bottom panel to illustrate the full range of isochrones used to derive photometric metallicities). The underlying black CMD in each subpanel is the one from the central pointing of Dw3 (upper left subpanel). \emph{Lower panels.} Metallicity distribution functions for each pointing in the upper panels, derived by interpolating metallicity values for individual RGB stars between isochrones with fixed age and varying
  metallicity. The foreground/background contamination has been subtracted using the Dw3S parallel pointing as "field" (lower central panel); the MDFs are normalized to the total number of stars for which metallicities have been derived. The median metallicity and the metallicity dispersion from the best-fitting Gaussian (red line) are reported for each field, as is the number of RGB stars included in the MDF. The underlying gray MDF in each subpanel is the one from the central pointing of Dw3 (upper left subpanel). For the central Dw3 pointing, we also report the MDFs with the respective median metallicities for stars within (red histogram) and beyond (blue histogram) $0.5 r_h$.}
\label{dw3_mdf}
\end{figure*}


\section{Comments on individual dwarfs} \label{individual}

\subsection{An ultra-diffuse galaxy and its possible companion: Dw1 and Dw2}

Dw1 and Dw2 were first presented in \citet{crnojevic14b}, and due to their projected separation of only 3~arcmin they appeared like a possible pair of Cen~A satellites. In light of their revised distances (Tab.~\ref{properties}), Dw2 seems to be located $\sim200$~kpc further away than Dw1, but the values are actually consistent within the error bars. While their distance from each other is unconstrained and we cannot confirm that they are physically bound, we stress that the probability of finding two dwarfs with such a small projected distance over our survey area is negligible.

Dw1 is a very intriguing galaxy on its own: it can be classified as an
ultra-diffuse galaxy \citep[given the definition of
][see also \citealt{Sandage84} for an early example of this type of galaxies]{vandokkum15}. The exquisite HST imaging allowed us to identify not
only a central star cluster with $M_V\sim-9$, but also a system of
three globular clusters within Dw1's half-light radius (with
$-7\lesssim M_V \lesssim-8$), all partially resolved into stars (a
thorough search for globular clusters around all of the PISCeS dwarf
discoveries will be presented in a future work). Spectroscopic
follow-up of the clusters has been obtained (Seth et al., in prep.),
and it will shed light on the properties of one of the closest
ultra-diffuse galaxies \citep[e.g.,][]{beasley16, amorisco18,
  bennet18, lim18}. Dw1's properties are overall reminiscent of the Fornax dwarf spheroidal in the Local Group.

For our analysis of Dw1's populations, we combine the Dw1 and Dw2 parallel WFC3 pointings to estimate the foregound/background contamination. 
Based on the non-negligible presence of AGB stars above the TRGB with $F814W_0\lesssim23.2$, Dw1 is likely to have formed stars until $\sim1-2$~Gyr ago. 
We note that a more detailed analysis of the fraction/ages/metallicities of AGB stars in Cen~A satellites is rendered difficult by the heavy contamination from foreground MW stars in the same CMD region: a proper decontamination will require near-infrared imaging, which allows for a clean separation of foreground sequences from stars at the distance of Cen~A \citep[see][]{crnojevic11a}.
Despite this extended star formation history, Dw1 does not presently contain any gas reservoir (see Tab.~\ref{properties}), as also confirmed by the absence of young stellar populations in its CMD. The RGB stellar density map from our Magellan imaging (the dwarf extends beyond the HST FoV) does not reveal any asymmetries, despite Dw1's non-negligible elongation ($\epsilon\sim0.2$), suggesting that this satellite has not been significantly perturbed by Cen~A. Its median metallicity is in line with that expected from the luminosity--metallicity relation derived for Local Group galaxies \citep[][their Fig.~12]{mcconnachie12}. We additionally derive the median metallicity within and beyond its half-light radius, obtaining [Fe/H]$\sim-0.99$ for the former and $\sim-1.07$ for the latter and thus revealing a mild metallicity gradient (uncertainties on the derived median metallicities are $\sim0.01$~dex). Gradients in dwarfs are routinely observed and they correlate well with galaxy luminosity \citep[e.g.,][]{leaman13}. More metal-rich (and likely younger) populations are more centrally concentrated, as is the case also for previously known Cen~A satellites \citep[see][and references therein]{crnojevic10}; this agrees well with the presence of intermediate-age AGB stars in Dw1.  

Dw2 contains predominantly old populations (Fig.~\ref{dwonetwo}), although a few candidate AGB stars ($F814W_0\lesssim23.5$) are observed above its TRGB within its half-light radius, pointing to a likely star formation episode between $\sim2$ and 4~Gyr ago. The foregound/background contamination for this satellite has been estimated from within the ACS FoV, which still contains the outskirts of Dw1. Despite its proximity to the latter, the RGB population of Dw2 shows a regular and rather circular shape, and its median metallicity is consistent with Dw2's absolute magnitude. Finally, the HIPASS upper limit on an HI reservoir in Dw1 only weakly constrains its gas richness.

\subsection{Dw4, Dw5, Dw6}

These three faint satellites of Cen~A, all located to its north, have a rather large galactocentric distance (85--125~kpc). Due to this, the CMD contamination for the equivalent area covered by each dwarf is negligible (see Figs.~\ref{dw4}, \ref{dw5}, \ref{dw6}). All three dwarfs have well-defined old RGB sequences; Dw4 and Dw6 additionally present a handful of bright sources at $F814W_0\lesssim23.6$ which may indicate a low level of star formation between 2 and 4~Gyr ago. Once again, HIPASS only returns weak upper limits on their neutral gas content. Among the three, Dw4 has the highest ellipticity ($\epsilon\sim0.3$), however there is no significant presence of debris that could point to tidal disruption for this dwarf. The faintest dwarf uncovered by PISCeS, Dw5, has an unperturbed appearance. Dw6's RGB map shows a very small overdensity in its outskirts to the West, which was already visible in the Magellan dataset; this may be connected to Dw6, or may simply represent a background fluctuation. Finally, Dw4 and Dw6 have a median metallicity that places them slightly above, but consistent with, the locus of Local Group dwarfs in the luminosity--metallicity relation.

\subsection{Dw7: a possible disruption?} 
Due to its proximity to Cen~A ($\sim$40 kpc in projection), the CMD of Dw7 is heavily contaminated by metal-rich populations (the ``field" CMD shown in Fig.~\ref{dw7} is extracted from the WFC3 parallel pointing). This metal-poor dwarf does not appear to host any young/intermediate-age populations, or detectable neutral gas.

From the RGB density map of Dw7, a small overdensity appears in the NW corner of our ACS pointing: the overdensity is $\sim8 \sigma$ above the mean stellar density in the lower half of the ACS pointing (which for convenience we call the ``primary field"). For reference, the number of Dw7 RGB stars within 1$r_h$ is $\sim60 \sigma$ above the ``primary field" level. Interestingly, at the same time the RGB density in the Dw7 WFC3 parallel pointing (or ``parallel field") is $\sim6 \sigma$ \emph{below} that in the ``primary field". We reconsider our ground-based RGB density map in the Dw7 region (upper right panel in Fig.~\ref{dw7}, where we overlay the location of our HST pointings): also within this catalog the RGB overdensity is visible and detected at a $\sim5 \sigma$ level. However, there is virtually no difference in number counts between the ``primary field" and ``parallel field" in the Magellan photometry. This may be attributed to the overall low number of RGB stars ($\sim40$ in Magellan vs the $\sim350$ in the ACS pointing, for the ``primary field"): the significantly deeper HST images allow a cleaner selection of RGB stars and unveil features that were not detected in the ground-based photometry. Given the small size of the HST pointings, it is not possible to investigate the presence of further overdensities around Dw7 from this dataset; a closer look at the ground-based map does not highlight unambiguous asymmetries/tails emanating from Dw7, as the area around this dwarf is heavily contaminated by Cen~A's halo stars and presents several randomly distributed overdensities. Finally, the radial RGB density profile of Dw7 does show an excess of sources at large galactocentric radii with respect to a simple exponential profile (the uncertainties on the structural parameters are indeed higher than for the other dwarfs, see Tab.~\ref{properties}). The lower stellar density found in the WFC3 pointing might imply that the ACS FoV contains a lingering low density stellar contribution from Dw7, even though our data are not deep enough to confirm an ongoing disruption.

\tabletypesize{\scriptsize}
\begin{deluxetable} {lccc}
\tablecolumns{4}
\tablecaption{Distances and metallicities along Dw3's tails.}
\tablehead{\colhead{Pointing} & \colhead{$(m-M)_0$ (mag)}  & \colhead{D (Mpc)} & \colhead{$[$Fe$/$H$]_{\rm med}$} }
\startdata
Dw3 - ACS & $27.94\pm0.09$  &  $3.88\pm0.16$ & $-0.61\pm0.01$ \\
Dw3 - WFC3 &  $28.29\pm0.08$  &  $4.54\pm0.17$ & $-0.66\pm0.01$ \\
Dw3S - ACS & $27.81\pm0.12$  &  $3.65\pm0.20$ & $-0.86\pm0.01$ \\
Dw3S - WFC3 & $28.10\pm0.25$  &  $4.16\pm0.51$ & $-0.65\pm0.01$\\
Dw3N - ACS &  $27.89\pm0.14$  &  $3.79\pm0.25$ & $-0.80\pm0.01$\tablenotemark{a}\\
Dw3N - WFC3 & $27.97\pm0.18$  &  $3.92\pm0.33$ & $-0.71\pm0.02$\tablenotemark{a}\\
\enddata
\tablenotetext{a}{The median metallicity has been derived excluding values with [Fe/H]$>-0.5$.}
\label{dw3_tails}
\end{deluxetable}

\subsection{The curious case of Dw3: population gradients along its tails} \label{sec:disrupting}

The largest coherent substructure in the halo of Cen~A is constituted by the disrupting Dw3 and its extended (more than 1~deg) tidal stream. The outer isophotes in the dwarf's remnant show an S-shape typical of tidal disruption \citep[see][]{crnojevic16}, although on a larger scale than the ACS FoV. Dw3 additionally hosts an elongated central star cluster (Seth et al., in prep.), which is partially resolved into stars in the HST imaging. 

The CMD for the inner region of Dw3's remnant is presented in Fig.~\ref{dw3}, and it shows a relatively broad RGB and a luminous AGB extending to colors even redder than that of the TRGB. The excess of AGB stars at $F814W_0\lesssim23.1$ with respect to the foreground/background rescaled field (which is extracted from the Dw3S parallel pointing, see below) suggests a prolonged star formation, possibly until $\sim$1--2~Gyr ago as indicated by the luminosity of the brightest AGB stars; later star formation is unlikely given the lack of blue sequences in Dw3's CMD. The HST dataset allows us to revise the distance of Dw3 and to place it at the same distance as Cen~A. As in \citet{crnojevic16}, we repeat the measurement of the RGB stars luminosity in the tails of Dw3 from our Megacam data \citep[see][]{crnojevic16}, adopting the new distance estimate, and we confirm that the original total magnitude of Dw3 (i.e., prior to tidal disruption) could have been as bright as $M_V\sim-15$. This can be further established from the median metallicity of Dw3's remnant, which would place it at a comparable absolute magnitude in Fig.~\ref{lummet}.

In Fig.~\ref{dw3_map} we show a zoom-in RGB stellar density map around the remnant of Dw3: the map is derived from our ground-based Magellan imaging, and overlaid are the positions of our follow-up HST pointings in this region (Dw1, Dw2, Dw3, Dw3S, and their relative parallel fields), along with one nearby pointing and its parallel, which we dub Dw3N, from program GO-12964 (PI Rejkuba, see \citealt{rejkuba14}; these fortuitously lay on top of Dw3's tidal tail). In the subsequent analysis, we will adopt the Dw3S parallel WFC3 pointing as the "field" for Dw3, since it is located off of the tail. In Fig.~\ref{dw3_map} we further report the CMDs for the HST pointings in this area, in order to look for possible gradients in the distances/stellar populations along Dw3's tails. 

We first compute the TRGB distance for each pointing (as described in Sect.~\ref{dist_sec}), which we report in Table~\ref{dw3_tails}, while the TRGB magnitudes are overplotted on each CMD. Overall, there is good agreement between the distances along the tails and that of Dw3's remnant within the uncertainties, and there is no strong indication of a distance gradient. The only discrepant pointing is the parallel of Dw3 (Dw3-WFC3), which seems to be $\sim600$~kpc more distant: given that the Dw3S pointing, at a larger Dw3-centric distance, has a distance consistent with Dw3, we suspect that the Dw3-WFC3 distance might be due to a combination of small number statistics and small-scale inhomogeneities in the stream populations. For the rest of this analysis, we will adopt Dw3's nominal distance for all the pointings. 

In the bottom of Fig.~\ref{dw3_map} we also present a RGB radial density profile as a function of distance from Dw3: also here, Dw3-WFC3 seems to contain fewer stars than expected from the overall profile. It is worth noting that the orientation and the shape of Dw3's stream are hard to reconcile with tidal disruption by Cen~A, however no other nearby galaxy could be responsible for its disruption \citep[see also][]{crnojevic16}. What is even more noteworthy is that the stellar density increases again beyond a galactocentric radius of $\sim0.3$~deg (i.e., for the pointings along the northern portion of the stream), instead of continuously decreasing as a function of radius. A similar result using the same HST data was  found by \citet[][their pointing F6]{rejkuba14}. We investigate this further by inspecting the CMDs of Dw3N and Dw3N-WFC3: they both seem to have an excess of sources with redder colors than the main RGB locus and at magnitudes fainter than the TRGB, when compared to all the other pointings. We perform the following test: we statistically subtract sources of the Dw3S CMD from the Dw3N CMD, given their comparable Dw3-centric distance. We plot the resulting CMD distribution of one realization of the subtraction in the bottom of Fig.~\ref{dw3_map}: the residual sequences resemble a more distant RGB with a possible well-populated luminous AGB extending to red colors. We derive a TRGB value of $\sim25.4$ for this putative stellar population, which would place it at a distance of $\sim7.6$~Mpc. The overdensity along the northern portion of Dw3's stream is clearly visible from our RGB density map, and the overdensity in the CMD space is confirmed by our test: while these are undeniable, the presence of a possible background object at such a large distance is puzzling. Its spatial properties cannot be investigated further (the Magellan dataset is too foreground/background contaminated), and a search of the NASA Extragalactic Database (NED) does not reveal any other background galaxies in this area. Even if the observed features were truly part of a background overdensity, it is impossible for this to have been interacting with Dw3 given their relative distances. 

To summarize, the northern portion of Dw3's stream presents an overdensity with respect to the expected radial decline (as evinced from the southern portion of the stream; Fig.~\ref{dw3_map}), which is possibly due to a distant, resolved background object. Next we will discuss the implications on the derived MDF of Dw3N.

The shapes of the RGB sequence in the Dw3 pointings differ significantly from each other (Fig.~\ref{dw3_mdf}, upper panels), suggesting a varying metallicity content.
We derive photometric metallicities as described above, and statistically subtract the ``field" contribution from the MDFs as derived from the Dw3S-WFC3 pointing, which is dominated by Cen~A's halo stars. The resulting MDFs and median metallicities are presented in the bottom panels of Fig.~\ref{dw3_mdf}; for the central Dw3 pointing, we additionally compute the MDFs within and beyond $0.5 r_h$. In Fig.~\ref{dw3_map} we also plot the median metallicities as a function of radius (with the relative 50--percentile metallicity intervals). 
A gradient of $-0.03$~dex/kpc is present within the central regions of the disrupting dwarf; along the southern portion of the stream the median metallicity decreases significantly ($-0.25$~dex over $\sim17$~kpc) with respect to Dw3's remnant, as also evidenced by the varying MDF shape between Dw3 and Dw3S (Fig.~\ref{dw3_mdf}). If we extrapolated the inner gradient to the outer regions of Dw3, the Dw3S populations would have originally come from a distance of $\sim8$~kpc from the center, rather than the presently observed $\sim17$~kpc. To justify this large Dw3-centric distance, either Dw3 has a past as an extremely diffuse galaxy, or its pre-disruption metallicity gradient must have been steeper than the one currently observed. The MDFs for the Dw3N pointings seem to contain a metal-rich population, stemming from the red sources identified in their CMDs and discussed above; this metal-rich ``tail" is not observed in the Dw3S MDFs. We thus assume these to be residual contaminants after our standard ``field" subtraction, and we compute the median metallicities for the Dw3N pointings excluding all metallicities [Fe/H]$>-0.5$. The resulting values are still higher than that expected from the gradient found along the southern tail portion (see the metallicity profile in Fig.~\ref{dw3_map}), because the distribution of contaminating sources in the CMD is indeed not limited to the reddest colors. 

We additionally note that the shape of the Dw3 MDFs is not well approximated by a Gaussian: this is not surprising and already observed in Local Group dwarfs for which spectroscopic MDFs have been derived \citep{kirby13}. In particular, the MDF fall-off on the metal-rich end can be explained by the effects of supernova explosions and stellar winds on the evolution of the dwarf \citep[see][and references therein]{crnojevic10}. We plan a more in-depth analysis of Dw3's stellar populations via spectroscopy obtained with VLT/VIMOS (Toloba et al., in prep.).
Finally, the difference in the MDFs of the Dw3 tails and of the "field" pointing (Fig.~\ref{dw3_mdf}) demonstrate how the material stripped off of this relatively massive dwarf could not be the primary source of populations in the Cen~A halo at these distances, since the latter peaks at significantly more metal-rich values (in accordance with the predictions of \citealt{dsouza17}; see also \citealt{rejkuba11, rejkuba14}).


\tabletypesize{\scriptsize}
\begin{deluxetable*} {lcccccc}
\tablecolumns{7}
\tablecaption{Absolute luminosities of Cen~A group members within 300~kpc of Cen~A ; those located beyond 150~kpc are below the horizontal line.}
\tablehead{\colhead{Galaxy} & \colhead{RA (h:m:s)} & \colhead{Dec (d:m:s)} & \colhead{Type} & \colhead{$M_V$ (mag)} & \colhead{D$_{\rm CenA, proj}$ (kpc)} & \colhead{Ref\tablenotemark{a}} }
\startdata
Dw11 & 13:17:49.2 & $-42$:55:37 & dSph? & $-9.4$ & 91& 3 \\
Dw5 & 13:19:52.4 & $-41$:59:41 & dSph & $-8.2$  & 94 & 3\\
KK196 & 13:21:47.1 & $-45$:03:48 & dSph & $-12.5$ & 139 & 1\\
KK197 & 13:22:01.8 & $-42$:32:08 & dSph & $-12.6$ & 51 & 1\\
$[$KK$2000]55$ & 13:22:12.4 & $-42$:43:51 & dSph & $-12.4$ & 43 & 1\\
Dw4 & 13:23:02.6 & $-41$:47:09 & dSph & $-9.9$  & 85 & 3\\
Dw10 & 13:24:32.9 & $-44$:44:07 & dSph? & $-7.8$  & 112& 3\\
NGC5128 & 13:25:27.6 & $-43$:01:09 & E & $-21.0$ & 0 & 1\\
Dw6 & 13:25:57.3 & $-41$:05:37 & dSph & $-9.1$  & 125 & 3\\
Dw7 & 13:26:28.6 & $-43$:33:23 & dSph & $-9.9$  & 37 & 3\\
ESO324-024 & 13:27:37.4 & $-41$:28:50 & dIrr & $-15.5$  & 103 & 2\\
Dw2 & 13:29:57.4 & $-41$:52:24 & dSph & $-9.7$  & 92 & 3\\
Dw1 & 13:30:14.3 & $-41$:53:35 & dSph & $-13.8$ & 93 & 3\\
Dw3 & 13:30:20.4 & $-42$:11:30 & dSph? & $-13.1$ & 79 & 3\\
Dw9 & 13:33:01.5 & $-42$:31:49 & dSph? & $-9.1$  & 95 & 3\\
Dw8 & 13:33:34.1 & $-41$:36:29 & dSph? & $-9.7$  & 133 & 3\\
NGC5237 & 13:37:38.9 & $-42$:50:51 & dSph & $-15.3$ & 145 & 1\\
\hline
KK189 & 13:12:45.0 & $-41$:49:55 & dSph & $-11.2$  & 170 & 1\\
ESO269-066 & 13:13:09.2 & $-44$:53:24 & dE & $-14.1$  & 188 & 1\\
KK203 & 13:27:28.1 & $-45$:21:09 & dSph & $-10.5$ & 153 & 2\\
ESO270-017 & 13:34:47.3 & $-45$:32:51 & S & $-17.1$  & 196 & 1\\
$[$KK$2000]57$ & 13:41:38.1 & $-42$:34:55 & dSph & $-10.6$  & 194 & 1\\
KK211 & 13:42:05.6 & $-45$:12:18 & dE & $-12.0$  & 240 & 1\\
KK213 & 13:43:35.8 & $-43$:46:09 & dSph & $-10.0$  & 219 & 1\\
ESO325-011 & 13:45:00.8 & $-41$:51:32 & dIrr & $-11.9$  & 246 & 1\\
\enddata
\tablenotetext{a}{References: 1=\citet{sharina08}, updated with latest \citet{kara13} distance measurements; 2=$M_B$ value from \citet{kara13}, from which $M_V$ is estimated as $M_B-0.31$ (see text for details); 3=this work.}
\label{lumfun_tab}
\end{deluxetable*}


\section{CenA satellite luminosity function} \label{lf}

Determining the faint end slope of the galaxy LF is crucial to constrain the physics governing galaxy formation and evolution at the smallest scales, and to understand the relation between stellar content and dark matter halo in dwarf galaxies. 
The ``missing satellite" problem around the MW implies a shallower LF slope ($\sim-1.2$; e.g., \citealt{koposov08}) than that predicted for the mass function of dark matter subhalos ($\sim-2.0$; e.g., \citealt{trentham02}). Several possible explanations (in terms of both observational incompleteness and theoretical modeling) have been put forward in the past decade to address this issue \citep[e.g.,][]{tollerud08,Brooks13,hargis14,Sawala16,Wetzel16,Garrison17,Kim17}: the general consensus is that the incorporation of prescriptions for feedback, star formation efficiency and reionization into cosmological simulations can help reconcile the observed LF slope with theoretical predictions \citep[see the recent review by][]{Bullock17}. However, the question of whether or not such models are tuned for the MW LF and whether or not they are capable of reproducing other systems remains.

The faint end of the satellite LF of further systems beyond the Local Group is key to understand the typical LF slope and its scatter from system to system. 

Beyond the Local Group, the measurement of the LF is complicated by two factors: the detection limits for satellites are significantly brighter and quickly fade with distance, and assessing the membership of candidate satellite galaxies with distance/velocity measurements is often prohibitively expensive. Perhaps not surprisingly, contrasting results have been derived for the LF in galaxy cluster and field environments, pointing to a possible dependence on environmental density \citep[e.g., see][and references therein]{ferrarese16}. Here, we focus on a sample of nearby ($<10$~Mpc) groups of galaxies, for which satellite memberships have been confirmed. With the aim to obtain as fair a comparison as possible in a range of group environments, we choose to perform an area-limited comparison.

We compile the cumulative luminosity functions (CLFs) for the Local Group and for nearby groups of galaxies with satellites confirmed via distance measurements. For the MW, we adopt the updated online 2015 version of the \citet{mcconnachie12} compilation; since the limiting magnitude for known dwarfs around the MW is significantly fainter than for all other groups, we only consider objects with $M_V<-5$, thus practically excluding all the recent extremely faint discoveries, e.g., from the DES (except for Eridanus~II which has $M_V=-7.1$, see \citealt{eriII}). For M31, we combine the catalogues presented in \citet{martin16} and \citet{mcconnachie18}. Arguably, M31 and its subgroup is the environment for which the LF is best constrained to date, both in terms of brightness limits, of spatial coverage, and of detection completeness: satellites as faint as $M_V\sim-6$ have been discovered out to a radius of approximately 150~kpc (M31's virial radius is estimated to be $\sim300$~kpc) over the course of the past decade \citep[see][and references therein]{mcconnachie12}.  

For galaxies beyond the Local Group, our main sources are the Updated Nearby Galaxy Catalogue \citep[][from which we only select satellites with positive tidal indexes]{kara13} and EDD\footnote{http://edd.ifa.hawaii.edu/}, the Extragalactic Distance Database \citep{jacobs09}. For M81, we complement these entries with Table~3 from \citet{chiboucas13}, who performed a CFHT/MegaCam wide-field survey of M81 to search for faint satellites which were then confirmed as group members with HST follow-up imaging \citep[see][for details]{chiboucas09,chiboucas13}. From this sample, we exclude possible tidal dwarfs. The M81 reported magnitudes are in $r$-band, $M_r$, which we convert to $M_V$ adopting the empirical relation derived from our Magellan dataset for Cen~A dwarfs ($M_V\sim M_r+0.4$). 
We also note that, for their faintest M81 satellite d0944+69, \citet{chiboucas13} report $M_I$ and $M_r$ that differ by $\sim2$~mag, thus the faintest datapoint of the M81 CLF is highly uncertain. For M101 we also adopt the \citet{kara13} catalogue, to which we add the three faint satellites discovered by Dragonfly (and later confirmed via distance measurements; see \citealt{merritt14, Danieli17}). Three galaxies close to M101 (NGC~5474, NGC~5477, and UGC~9405) do not have direct TRGB distance measurements, but are considered as likely distant group members. The M94 spiral has been recently surveyed by \citet{Smercina18}, who added two faint satellites with TRGB distances to only two other likely distant group members (KK160 and IC3687), making this the most poorly populated environment of our sample. Finally, for Cen~A we complement the results from this paper with the catalog from \citet{kara13}: the listed $M_B$ magnitudes are transformed into $M_V$ by applying the conversion $M_V=M_B-0.31$. This relation is derived for a subsample of satellites for which $M_V$ values are reported in \citet{sharina08}, and updated by applying the latest distance measurements. We have compiled an updated table of the Cen~A satellites with projected distances $<300$~kpc, including their coordinates, projected distances and luminosities (Table \ref{lumfun_tab}).

\begin{figure}
 \centering
\includegraphics[width=8cm]{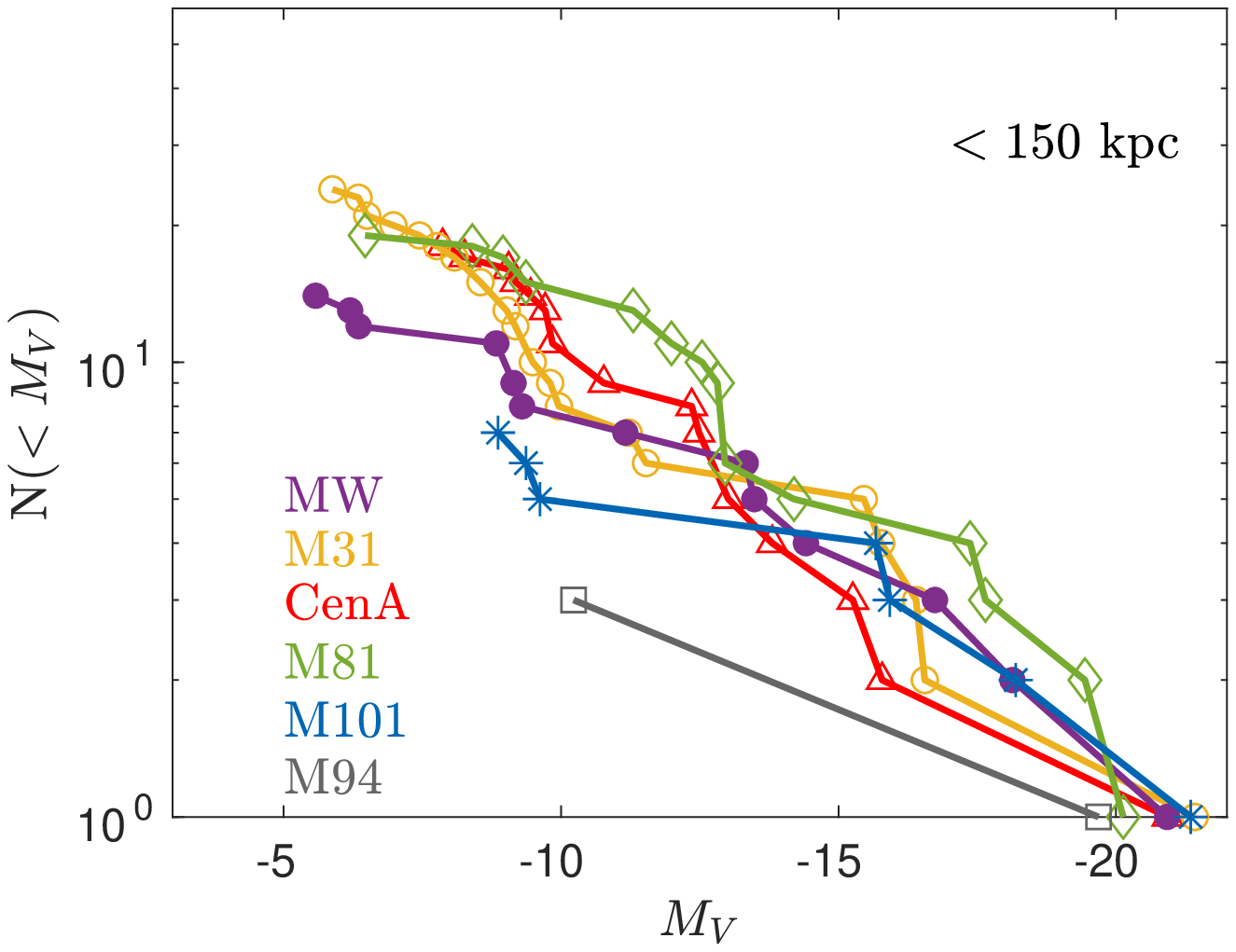}
\includegraphics[width=8cm]{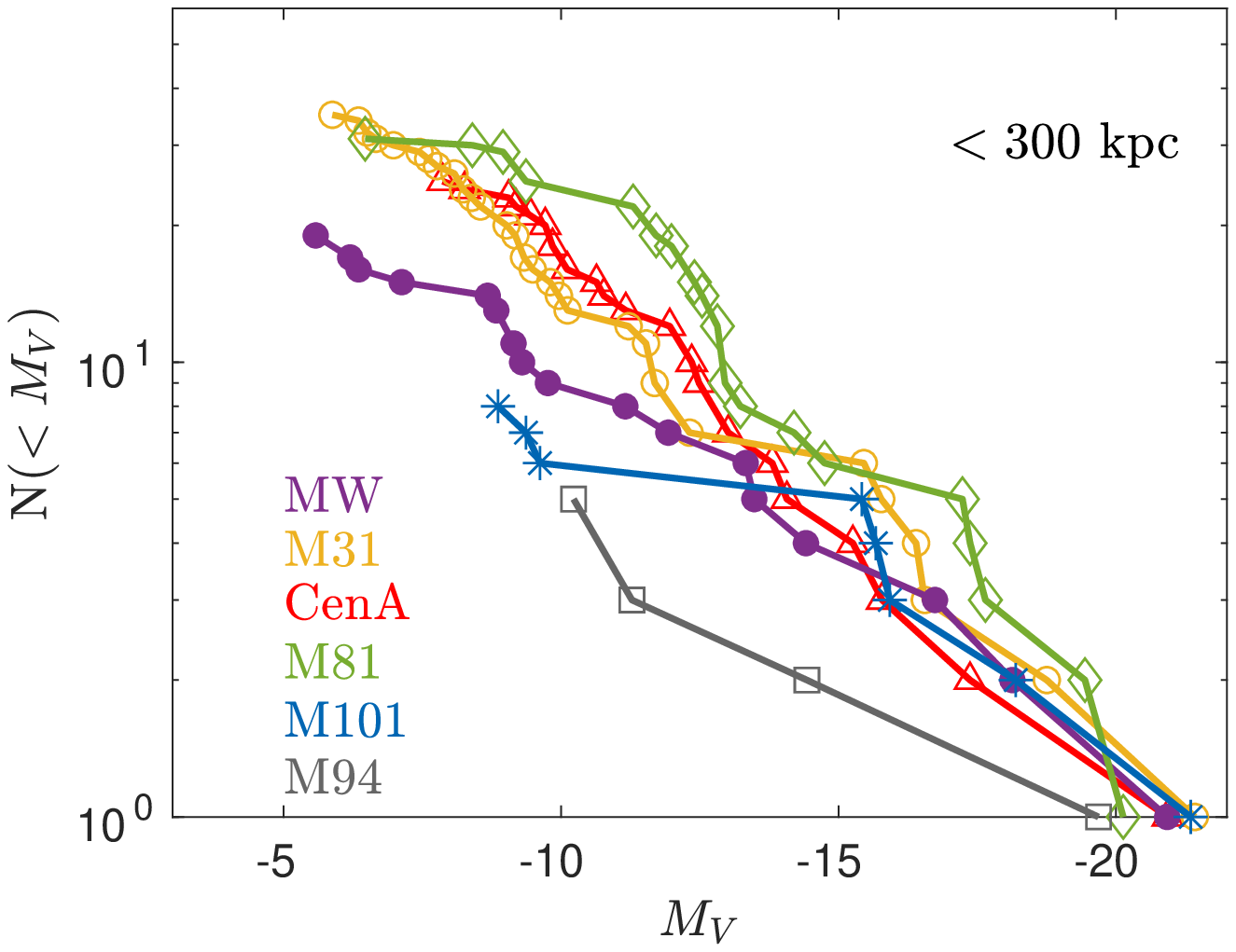}
\caption{Cumulative galaxy LFs for Cen~A (red triangles and line), the MW (purple dots), Andromeda (yellow circles), M81 (green diamonds), M101 (cyan asterisks), and M94 (gray squares). The \emph{upper panel} shows satellites within a projected radius (or 3D radius for the MW) of 150~kpc; the \emph{lower panel} includes objects within 300~kpc of each host (see text for details). For the latter sample, a cumulative Schechter function gives faint-end slopes $\alpha$ of: $-1.14^{+0.17}_{-0.16}$ for Cen~A, $-1.13^{+0.08}_{-0.08}$ for the MW, $-1.22^{+0.11}_{-0.10}$ for Andromeda, $-1.06^{+0.31}_{-0.12}$ for M81, $\sim -0.80$ for M101, and $\sim -1.16$ for M94 (the latter two are poorly constrained).
}
\label{lumfun}
\end{figure}

The area coverage of the different surveys we consider is not easy to quantify in light of the underlying distribution of satellite galaxies, and as mentioned before, we restrict the derived CLFs by area. The PISCeS survey has been designed to cover 150~kpc in radius around its target galaxies Cen~A and Sculptor, offering the advantage of a relatively straightforward comparison to the PAndAS survey of M31. In the lower panel of Fig.~\ref{lumfun}, we only consider satellites with distances $<150$~kpc from the respective host; such distances are necessarily projected, except for the MW where 3D distances are adopted. In the upper panel of Fig.~\ref{lumfun}, we additionally draw the CLF for group members found within the virial radius of each host: the latter is an uncertain quantity, and we assume it to be $\sim300$~kpc given the comparable luminosities of our sample of galaxies \citep[e.g.,][]{klypin02}. A few caveats are worth mentioning: i) MW surveys inevitably suffer from incompleteness effects (mainly due to incomplete spatial coverage, especially in the direction of the Galaxy plane), which may underestimate the number of faint satellites by a factor of $\sim3$ \citep[e.g.,][]{tollerud08, hargis14}; ii) for Cen~A, a number of candidate satellites have been discovered as unresolved low surface brightness objects in a DECam imaging survey of $\sim500$~deg$^2$ around Cen~A: \citet{mueller17} present $\sim40$ new candidate satellites, of which 13 are located within its virial radius (but none within the PISCeS footprint). These candidates await distance measurements to be confirmed as group members and they are not included in the CLF, which could thus be steeper than the one we construct in the magnitude range $-8\lesssim M_V \lesssim -12$; iii) for M101, a similar search for low surface brightness galaxies has been presented in \citet{bennet17}, thus the CLF for this group might also be a lower limit at its faint end ($M_V>-10$); iv) a deep search for faint satellites has not been performed beyond the innermost 150~kpc for M94, thus the CLF within the virial radius for this host is likely a lower limit for $M_V\gtrsim-11$.

Although artificial galaxy tests have not been performed yet for PISCeS, we can estimate our dwarf detection incompleteness by considering our discoveries: our limiting absolute magnitude and surface brightness are $M_V\sim-8$ and $\sim26.5$~mag/arcsec$^2$, respectively. The major factor impacting our ability to find new dwarfs is the highly varying seeing conditions under which our ground-based survey was performed ($\sim0.5$--$1.0$"). The PISCeS dwarfs with the lowest central surface brightness values were discovered in fields with seeing in at least one of the bands of 0.65" or better; in terms of absolute magnitude, the same seeing limit allows us to uncover objects with $M_V\sim-9$, while the $M_V\sim-8$ satellites were found under slightly better seeing conditions ($\sim0.6"$); the dwarf discovered under the worst seeing conditions ($\sim0.8"$) has an absolute magnitude of $M_V\sim-10$. Among the PISCeS pointings, $\sim10/35/50\%$ have seeing worse than 0.8/0.65/0.6", respectively: we thus assess our completeness to be around $\sim90/70/50\%$ for absolute magnitudes of $M_V\sim-10/-9/-8$, respectively. The incompleteness limits will additionally depend on the dwarfs' stellar concentration, on their distribution around Cen~A, and on spatial coverage (e.g., galaxies not detected because of bright foreground stars); see \citet{chiboucas09,Smercina18}, but these factors will not significantly alter our main conclusions. Our faintest discoveries all have half-light radii in the range 0.2--0.6~arcmin: even if we had not resolved them into stars, the compact size would have likely allowed us to identify them visually as unresolved low surface brightness objects (as for our unresolved candidates that turned out to lie in the background, see Sect.~\ref{confirm_sec}). The regime we are least sensitive to is the one at low surface brightness and large half-light radii, with an extremely faint unresolved component: the pointings containing Dw1 and Dw3 (our brightest and most diffuse discoveries) had excellent seeing conditions ($<0.55"$ in both bands), and PISCeS is thus not suited to uncovering faint ($M_V\gtrsim-13$) galaxies with such properties, assuming that they exist (see Fig.~\ref{scaling}). With these numbers in mind, in the magnitude range $-10<M_V<-8$ we might be missing 5--10 galaxies: adding those to Cen~A's CLF would not alter its slope within the uncertainties. As mentioned earlier, there are 13 additional unconfirmed candidates with galactocentric distances between 150 and 300~kpc from \citet{mueller17} in the magnitude range $-12<M_V<-8$: these were discovered from integrated light and a fraction of them could be background objects (none of our unresolved candidates was confirmed as a real Cen~A satellite, Sect.~\ref{confirm_sec}), thus we do not consider them further. That said, the \citet{mueller17} study did recover several of the dwarfs originally found by PISCeS \citep{crnojevic16}, so there may yet be true CenA satellites to be confirmed at such large radii.

We fit a Schechter function to each CLF within the virial radius:

\begin{equation}
 N(<M) = \phi_*\gamma[\alpha+1,10^{0.4(M_*-M)}],
\end{equation}

\noindent and we report the best-fit $\alpha$ values in the caption of Fig.~\ref{lumfun} ($\phi_*$ and $M_*$ are not well constrained, but they do not significantly affect the slope $\alpha$; see also \citealt{chiboucas13,park17}). The slopes are consistent among the different groups, as well as with previous literature results. In both galactocentric distance ranges, we observe a large scatter in the CLFs at fixed magnitude, but there is not an obvious link between the CLF faint end slope and the luminosity of the giant host, which is similar among the considered groups. 

\section{Discussion and conclusions} \label{disc}

We have presented HST follow-up imaging of seven newly discovered dwarfs in our PISCeS panoramic survey of Cen~A; four additional candidate satellites were found not to be group members as evidenced by a lack of resolved populations in the HST images. We also discovered two more dwarfs in our ground-based 2017 Magellan imaging, bringing the total number of new PISCeS Cen~A satellites to eleven (also including the two objects presented in \citealt{crnojevic16} for which HST data were not obtained). Before PISCeS, thirteen satellites were known within its estimated virial radius ($\sim300$~kpc), of which five were located within 150~kpc of the giant elliptical: our sample thus almost doubles Cen~A's satellite population.

The exquisitely deep HST imaging allowed us to derive updated values for the distances, luminosities, structural parameters and photometric metallicities for the target Cen~A dwarfs. With respect to the discovery paper \citep{crnojevic14a}, Dw1's absolute magnitude and surface brightness are revised to be significantly brighter, placing this dwarf into the ultra-diffuse category; Dw3, which is heavily disrupting, can also be considered an ultra-diffuse galaxy at the present time. Both Dw1 and Dw3 are found to host metallicity gradients; the tidal tails of Dw3 are more metal-poor than the surrounding populations in Cen~A's halo, indicating that satellites of similar luminosity likely did not contribute to the build-up of its outer halo. All the other discoveries from PISCeS are relatively faint and compact objects.
The eleven new PISCeS dwarfs are all predominantly old with no signs of recent ($\lesssim1$~Gyr) star formation: within 150~kpc of Cen~A, the majority of satellites are not currently forming stars, with the exception of ESO324-024, NGC~5237 and KK196 (all previously known). This will depend on their absolute luminosity (the most luminous satellites have been able to form stars until the present day), as well as on their group infall time (indeed they are among the most distant dwarfs from Cen~A, at least in projection). The PISCeS dwarfs extend the previously known satellite population $\sim2$~mag fainter in both absolute magnitude and central surface brightness (Fig.~\ref{scaling}): among the "classical" dwarfs, only one has $M_V\sim-10$, while nine out of our eleven new discoveries are fainter than this limit, and all of them have central surface brightness values fainter than the previous $\sim24.5$~mag/arcsec$^2$ limit. 
No ultra-faint dwarfs were uncovered by PISCeS, but several ultra-compact dwarf candidates are  being followed-up spectroscopically (Voggel et al., in prep.). Finally, the range in half-light radii is extended both to smaller and larger values with respect to the ``classical" dwarfs.

\citet{mueller16} conducted a thorough analysis of both confirmed and candidate dwarfs around Cen~A (including our PISCeS discoveries from \citealt{crnojevic16}) to investigate the two possible planes of satellites presented by \citet{tully15}, concluding that the presence of one single plane is more likely. Recently, \citet{mueller18} additionally presented evidence for a rotating plane of satellites around Cen~A: it will be interesting to collect kinematic data for the PISCeS dwarfs to investigate whether they belong to this whirling plane. We note that there is a visible asymmetry in the spatial distribution of PISCeS dwarfs, with eight out of eleven dwarfs located to the north of Cen~A's minor axis (coincident with its dust lane); curiously, five out of our eleven dwarfs are at a galactocentric distance of $\sim90$~kpc.

We investigated the CLF of Cen~A within 150 and 300~kpc (i.e., the estimated virial radius), and compared it to those of nearby groups with confirmed faint dwarf satellites (the MW, M31, M81, M101 and M94), spanning a range of host galaxy morphologies, and environments (from the relatively isolated M94 and M101, to the rich groups of M81 and Cen~A). While the derived faint-end slopes for the various groups are consistent within the sample and with previous literature work, the scatter in the CLFs is significant and does not correlate with the host galaxy mass. Recently, \citet{Smercina18} performed a similar study of the CLFs in the same nearby groups we consider in this work (except Cen~A), and showed that simulations cannot reproduce their observed scatter: the solution they put forward is a halo occupation model where the stellar mass-halo mass relation includes an increased scatter, suggestive of a highly stochastic galaxy formation efficiency in dark matter halos. Clearly, this topic deserves further attention from the theoretical standpoint, and dedicated simulations aimed at reproducing the observed CLFs beyond the Local Group are highly desirable.


\acknowledgments

The authors thank the referee, whose comments helped improve the presentation of our results.
DC wishes to kindly thank the hospitality of the Carnegie Observatories, 
where part of this work has been carried out. DC warmly thanks M. Rejkuba for
useful discussions. Support for this work was partly provided by NASA through grants number HST-GO-13856.001 and HST-GO-HST-GO-14259.001-A from the Space Telescope Science Institute, which is operated by AURA, Inc., under NASA contract NAS 5-26555.
DC acknowledges support from NSF grant AST-1814208; research by DJS is supported by NSF grants AST-1821967, 1821987, 1813708 and 1813466;
SP acknowledges support from NASA grant NNX14AF84G;
PG and ET acknowledge support from NSF grants AST-1010039 and AST-1412504; JDS acknowledges support from NSF grant AST-1412792; JS was supported by NSF grant AST-1514763 and a Packard Fellowship; KS acknowledges support from the Natural Sciences and Engineering Research Council of Canada (NSERC). 
This work was supported in part by National Science Foundation Grant No. 
PHYS-1066293 and the Aspen Center for Physics.
This paper uses data products produced by the OIR Telescope
Data Center, supported by the Smithsonian Astrophysical
Observatory.
This research has made use of the NASA/IPAC Extragalactic 
Database (NED) which is operated by the Jet Propulsion Laboratory, 
California Institute of Technology, under contract with the National 
Aeronautics and Space Administration. 


%

\vspace{5mm}
\facilities{Magellan Telescopes, Las Campanas Observatory, Chile (Megacam); Hubble Space Telescope}

\bibliographystyle{aasjournal}
\bibliography{biblio}

\clearpage




\end{document}